\documentclass[aps,prb,reprint,superscriptaddress,floatfix]{revtex4-2}

\usepackage[english]{babel}
\usepackage{graphicx}
\usepackage{amsmath,amssymb,amsthm,mathtools}
\usepackage{physics}
\usepackage{bm}
\usepackage{xcolor}
\usepackage{siunitx}
\usepackage{hyperref}
\hypersetup{
    colorlinks=true,
    citecolor=blue,
    linkcolor=red,
    urlcolor=blue
}

\newcommand*{\hham}{\hat{\mathcal{H}}}
\newcommand*{\Sop}{\hat{\mathbf{S}}}
\newcommand*{\sop}{\hat{\mathbf{s}}}
\newcommand{\ii}{\mathrm{i}}

\newcommand{\Gret}{\mathbf{G}^{R}}

\newcommand{\Gless}{\mathbf{G}^{<}}

\begin{document}

\title{Two Exceptional-Point Regimes in the Spinterface Model of Chirality-Induced Spin Selectivity}

\author{Seif Alwan}
\affiliation{Materials Theory, Department of Physics and Astronomy, Uppsala
University, Box 516, 75120 Uppsala, Sweden}

\date{\today}

\begin{abstract}
The spinterface model of chirality-induced spin selectivity supports two
algebraically distinct near-exceptional-point (EP) regimes, which cannot be
mixed. A \emph{gold-level} near-EP involves the two gold spin states under an
anti-Hermitian, electrode-mediated spin flip with spin-symmetric dissipative
broadening (Scheme~B). A \emph{chain-level} near-EP involves the collective
coalescence of the two molecular eigenstates with largest weight on the first
molecular site under a Hermitian on-site Rashba spin flip with spin-dependent
broadening (Scheme~A). Mixing the diagonal cancellation of one scheme with the
off-diagonal structure of the other yields a strictly negative discriminant and
admits no EP on any parameter path, making the gold/chain classification
exhaustive. At the gold near-EP the spin-flip coupling is chirality-selective,
converting achiral spin-valve transport into purely chiral transport across
well-defined crossover points. At the chain near-EP the Petermann factor grows
as a power law in chain length, with an exponent tunable by the Rashba
coherence length and non-monotonic in the Rashba strength, reflecting a balance
between coherent enhancement and progressive breakdown of the Scheme-A
condition. Both near-EPs are dressed versions of the ideal two-state EP, and
the ratio between observed and bare EP locations measures the delocalisation of
the coalescing pair, providing a fit-free diagnostic of the mechanism. These
results identify CISS junctions as a solid-state platform for non-Hermitian
spectral physics.
\end{abstract}

\maketitle

\section{Introduction}
\label{sec:intro}

The spinterface model of chirality-induced spin selectivity (CISS) has recently
been recently been formulated in a fully quantum form within the
non-equilibrium Green's function (NEGF) framework~\cite{Alwan2026QSpinterface}.
In that formulation, the exact antiunitary symmetry $H^*(\phi)=H(-\phi)$ of the
Hamiltonian implies a reciprocity relation
\begin{equation}
\bigl|\Gret_\phi[i,j]\bigr| = \bigl|\Gret_{-\phi}[j,i]\bigr|
\label{eq:reciprocity_intro}
\end{equation}
for the retarded Green's function. In a device that is left-right asymmetric in
the magnetic sense, as in the spinterface junction, where the magnetic moment
sits on the gold end and not on the molecular end~\cite{Alwan2026QSpinterface,
Alwan2021}, this relation does not force the transmission to be even in the
chirality angle $\phi$. The odd (chiral) part of the spin current,
$I_{S_z}^{\rm odd}(\phi)=\tfrac12[I_{S_z}(\phi)-I_{S_z}(-\phi)]$, is therefore
a genuine manifestation of reciprocity breaking, and it vanishes below a
threshold of four total sites because both a non-coplanar geometry (four
points) and two or more Rashba-active bonds are required for the breaking to
survive.

CISS remains a physically contested effect~\cite{Naaman2015, Evers22}.
Structural explanations, based on chirality-induced effective spin-orbit
fields, and interfacial explanations, based on spin-selective hybridisation at
a magnetic electrode, compete to explain the same body of experimental
observations. A diagnostic that is \emph{specific} to one mechanism, and that
is predicted to be absent under competing mechanisms, is therefore highly
desirable. Here we ask a question complementary to that of the
reciprocity-based formulation of Ref.~\cite{Alwan2026QSpinterface}: \emph{what
happens when the non-Hermitian structure that produces reciprocity breaking
becomes singular?} A singularity of the effective non-Hermitian Hamiltonian is
an exceptional point (EP), and EPs inherit the reciprocity-breaking structure
but amplify it: at an EP the forward and backward transmissions become not
merely unequal but \emph{structurally} unequal, with the coalescing
eigenvectors losing biorthogonal normalisation entirely. Because EPs are
singular points of the non-Hermitian structure, they leave signatures, these
are signature are a result of the coalescence of eigenvalues, divergence of the
Petermann factor, chirality-selective transport, that are quantitatively
different from what a Hermitian or weakly non-Hermitian competing mechanism
would produce.

Exceptional points are degeneracies of non-Hermitian systems at which two
eigenvalues \emph{and} their eigenvectors coalesce~\cite{ElGanainy2018,
Miri2019}. At an EP the spectrum develops a square-root branch point in the
complex energy plane, the biorthogonal overlap of the coalescing eigenvectors
vanishes, and the Petermann factor of the coalescing pair
diverges~\cite{Petermann1979,Siegman1989}. These features have made EPs the
paradigm of non-Hermitian spectral physics and the subject of intensive study
in optics, photonics, and cold atoms. Non-Hermitian concepts have recently
entered the CISS literature from several directions: the non-Hermitian skin
effect has been proposed as the origin of anomalous magnetoresistance in
chiral molecules~\cite{Zhao2025NHSE}; non-Hermitian exchange has been
identified as a unifying origin of CISS~\cite{Theiler2026SciAdv}; EPs
themselves have been shown to be enantiosensitive in open chiral
systems~\cite{Mayer2026EnantioEP} and have been realized in single-molecule
junctions~\cite{Yang2024SpiroEP}; and chiral phonons provide a further
channel via angular-momentum transfer to electronic
spins~\cite{Fransson2023ChiralPhonon,Juraschek2025ChiralPhonons}. In the
present paper the EPs we identify are \emph{near}-EPs: the two-level subspace
is dressed by virtual transitions to the rest of the spectrum, and the
resulting residual gap limits the Petermann divergence to a finite peak. The
finite value is a controlled diagnostic of the degree of dressing, not a
failure of the EP mechanism, and it is precisely this dressing that allows
the two-state predictions to be tested against the full NEGF calculation.

\begin{figure*}[htbp]
\centering
\includegraphics[width=\linewidth]{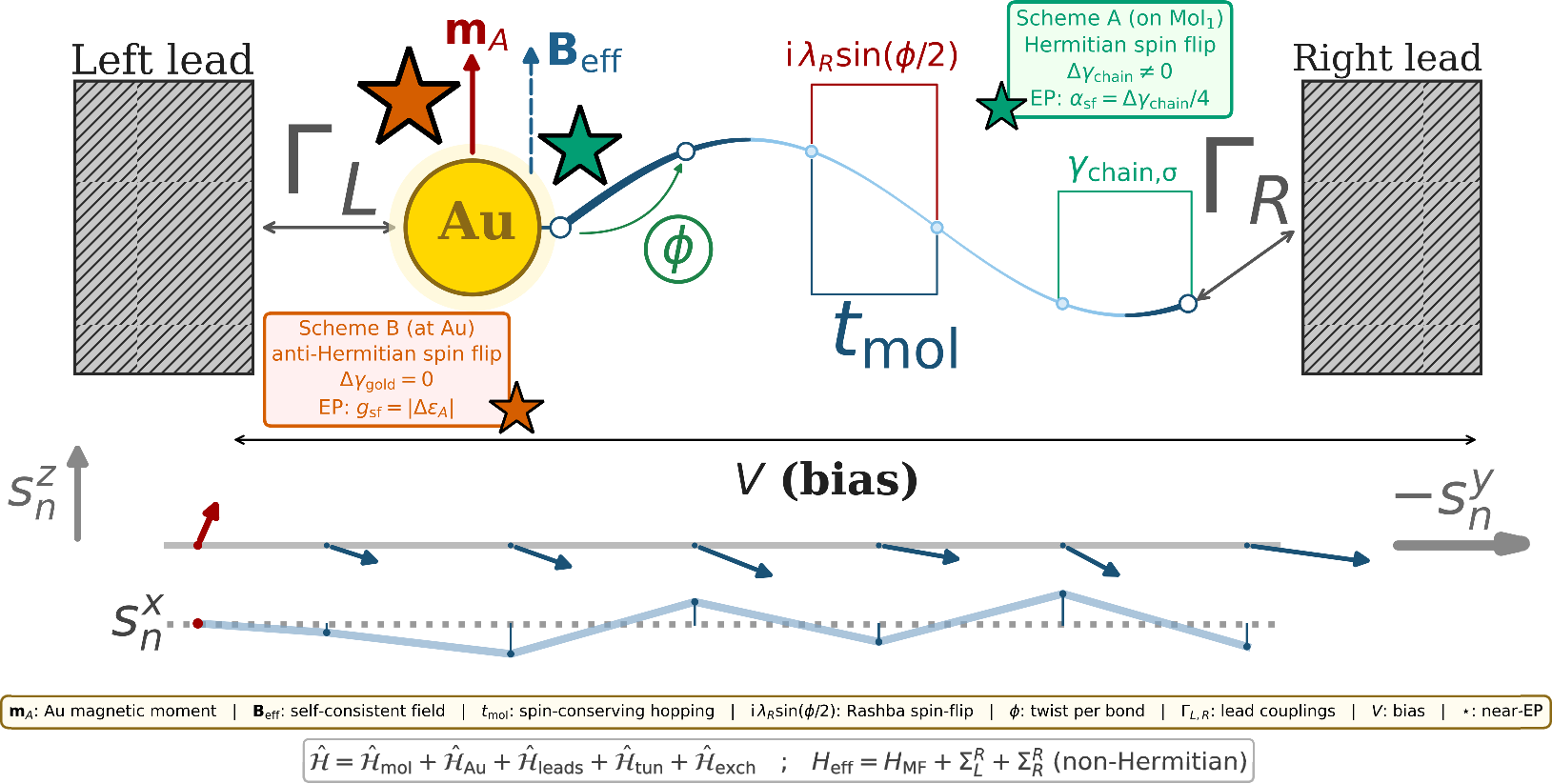}
\caption{%
Schematic of the spinterface model and its two exceptional-point
schemes. The junction comprises a gold atom (Au) with a local mean-field
moment $\mathbf{m}_A$, a chiral Rashba chain of $L$ sites with
nearest-neighbour hopping $t_{\rm mol}$ and Rashba spin-flip amplitude
$\ii\lambda_R\sin(\phi/2)$, and two non-magnetic leads coupled through
$\Gamma_{L,R}$ under an applied bias $V$. The two coloured EP stars
mark the two coalescing subspaces: red Scheme~B at the gold site
(anti-Hermitian spin flip, spin-symmetric leads, EP at
$g_{\rm sf}=|\Delta\varepsilon_A|$) and green Scheme~A at the first
molecular site (Hermitian spin flip, spin-dependent chain broadening
$\gamma_{\rm chain,\sigma}$, EP at
$\alpha_{\rm sf}=\Delta\gamma_{\rm chain}/4$). The dark-red box states
the algebraic obstruction to mixing the two schemes. The two lower rows
show the local spin texture $\langle\mathbf{s}_n\rangle$: upper row
$(s_n^y,s_n^z)$ arrows, lower row $s_n^x$ lollipops on a dashed
zero-line. Parameters: $L=6$, $\Delta\gamma=0.4$~eV,
$\varepsilon_A=1.5$~eV, $\phi=60^\circ$, $T=300$~K.}
\label{fig:EPschematic}
\end{figure*}

\paragraph{The two EP regimes.}
The spinterface model supports two EP regimes that differ not only in which
pair of eigenvalues coalesces but in the \emph{structure} of the effective
non-Hermitian coupling. A \emph{gold-level} near-EP involves the two gold
spin states, with spin-symmetric dissipative broadening and an anti-Hermitian,
electrode-mediated spin flip (Scheme~B). A \emph{chain-level} near-EP involves
the two molecular eigenstates with the largest amplitude on the first
molecular site, with spin-dependent dissipative broadening and a Hermitian,
on-site Rashba spin flip (Scheme~A). The two schemes cannot be mixed:
combining the diagonal cancellation of one scheme with the off-diagonal
structure of the other gives a strictly negative discriminant and admits no
EP on any parameter path. This algebraic obstruction makes the gold/chain
classification exhaustive: every near-EP the model admits is of one scheme or
the other. The two mechanisms, the EP conditions, and the algebraic
obstruction are derived in Sec.~\ref{sec:two_schemes} and summarised
schematically in Fig.~\ref{fig:EPschematic}.

\paragraph{Preview of the signatures.}
At both near-EPs the gap minimum and the Petermann maximum coincide to within
the grid resolution, a non-trivial consistency check that distinguishes a
genuine near-EP from an accidental spectral degeneracy. At the gold near-EP,
the anti-Hermitian spin flip is \emph{chirality-selective}: as it is tuned, it
coherently mixes the two gold spin states and progressively converts achiral
spin-valve transport into chiral CISS transport. The even and odd parts of the
spin current cross, and the achiral contribution can be driven to vanish
entirely while the chiral part remains finite. Because a competing
non-spinterface mechanism would not produce an achiral part that vanishes as
a function of an external spin-flip coupling, the observation or absence of
this extinction point is a direct experimental test of the spinterface
mechanism. At the chain near-EP, the Petermann factor of the coalescing pair
grows as a power law in chain length, and the exponent is tunable by the
Rashba coherence length. The dependence is \emph{non-monotonic}: the exponent
peaks at intermediate Rashba strength and becomes negative at strong Rashba,
reflecting a balance between coherent enhancement of the coalescence and
progressive breakdown of the Scheme-A condition. Both near-EPs are
quantitative, dressed versions of the ideal two-state EP: the ratio between
the observed and bare EP location measures the delocalisation of the
coalescing pair, providing a diagnostic of the mechanism that does not
require fitting to a specific model.

\paragraph{Paper structure.}
Section~\ref{sec:model} reviews the spinterface Hamiltonian and its NEGF
treatment, states the two EP schemes in full, and derives the
Petermann-factor scaling near a generic EP. Section~\ref{sec:gold} presents
the gold-level near-EP results. Section~\ref{sec:chain} presents the
chain-level near-EP results, including the full Rashba-strength landscape and
the collective scaling analysis. Section~\ref{sec:conclusion} closes with a
summary and outlook.

\section{Model and Methods}
\label{sec:model}

\subsection{Spinterface Hamiltonian and reciprocity}
\label{sec:spinterface_H}

We work with the spinterface model of Refs.~\cite{Alwan2021,
Alwan2026QSpinterface, Sarkar2025}. Throughout, $d_{n\sigma}$
($d_{n\sigma}^\dagger$) annihilates (creates) an electron on molecular site
$n\in\{1,\ldots,L\}$ with spin $\sigma\in\{\uparrow,\downarrow\}$, and
$c_{\alpha k\sigma}$ ($c_{\alpha k\sigma}^\dagger$) annihilates (creates) an
electron in lead $\alpha\in\{L,R\}$ at momentum $k$ with spin $\sigma$. The
gold site is labelled $A$ and hosts the fermionic operators $d_{A\sigma}$; the
last molecular site is labelled $B\equiv M_L$ for brevity. The Pauli matrices
are $\bm\sigma=(\sigma^x,\sigma^y,\sigma^z)$ and the on-site number operators
are $\hat n_{A\sigma}=d_{A\sigma}^\dagger d_{A\sigma}$.
The total Hamiltonian is
\begin{equation}
  \hham = \hham_{\rm mol} + \hham_{\rm Au} + \hham_{\rm hop} + \hham_{\rm leads} + \hham_{\rm tun} + \hham_{\rm exch},
\label{eq:Htot}
\end{equation}
with the molecular part
\begin{equation}
\hham_{\rm mol}
=
\sum_{n,\sigma} \varepsilon_n d_{n\sigma}^\dagger d_{n\sigma}
-
\sum_{\langle n,n'\rangle,\sigma} t_{nn'}\, d_{n\sigma}^\dagger d_{n'\sigma}
+ \text{h.c.}.
\label{eq:Hmol}
\end{equation}
Here $\varepsilon_n$ is the on-site energy of molecular site $n$, taken as
$\varepsilon_n\equiv\varepsilon_{\rm mol}$ for all molecular sites in the
numerical work below; $t_{nn'}$ is the nearest-neighbour hopping amplitude,
taken as $t_{nn'}\equiv t_{\rm mol}$ for nearest-neighbour pairs and zero
otherwise; and $\langle n,n'\rangle$ denotes a sum over nearest-neighbour
pairs.
While the gold-molecule hopping is given by
\begin{equation}
\hham_{\rm hop}
=
-t_{AB}\sum_\sigma\left(d_{A\sigma}^\dagger d_{1\sigma}+\text{h.c.}\right).
\label{eq:Hhop}
\end{equation}
The gold atom hosting a local moment $\Sop_A$,
\begin{equation}
\hham_{\rm Au} = \sum_{\sigma} \varepsilon_{A\sigma} d_{A\sigma}^\dagger d_{A\sigma},
\label{eq:HAu}
\end{equation}
The gold on-site energy is spin-dependent, with
\begin{equation}
\varepsilon_{A\sigma} = \varepsilon_A + \sigma\left(\tfrac{J_{el}}{2}+J'_p\right)m_z,
\label{eq:eps_A_sigma}
\end{equation}
where $\varepsilon_A$ is the bare on-site energy and $m_z$ is the
self-consistent moment defined below; the factor $\sigma=+1$ ($-1$) applies
to spin $\uparrow$ ($\downarrow$).
The leads
\begin{equation}
\hham_{\rm leads} = \sum_{\alpha\in\{L,R\}}\sum_{k,\sigma}\varepsilon_{\alpha
k\sigma} c_{\alpha k\sigma}^\dagger c_{\alpha k\sigma},
\label{eq:Hleads}
\end{equation}
the tunnelling
\begin{equation}
\hham_{\rm tun}
=
\sum_{k,\sigma}\left(V_L c_{Lk\sigma}^\dagger d_{A\sigma}+\text{h.c.}\right)
+
\sum_{k,\sigma}\left(V_R c_{Rk\sigma}^\dagger d_{B\sigma}+\text{h.c.}\right),
\label{eq:Htun}
\end{equation}
In Eq.~\eqref{eq:Hleads}, $\varepsilon_{\alpha k\sigma}$ is the dispersion
of lead $\alpha$; the leads are non-magnetic, so
$\varepsilon_{\alpha k\uparrow}=\varepsilon_{\alpha k\downarrow}$. In
Eq.~\eqref{eq:Htun}, $V_L$ and $V_R$ are the tunnelling matrix elements
between the left lead and the gold site, and between the right lead and
the last molecular site $B$, respectively; both are taken to be
momentum-independent in the wide-band limit used below.

The exchange/spin-orbit interaction
\begin{equation}
  \begin{split}
\hham_{\rm exch}
&=
J_{el}\,\Sop_A\cdot\sop_A
+
J_{p}'\,\hat S_A^z(\hat n_{A\uparrow}-\hat n_{A\downarrow})\\
&+
K\,\hat S_A^z\sum_\sigma\left(d_{A\sigma}^\dagger d_{B\sigma}+\text{h.c.}\right),
\label{eq:Hexch}
\end{split}
\end{equation}
where
$\sop_A=\tfrac12\sum_{\sigma\sigma'}d_{A\sigma}^\dagger\bm\sigma_{\sigma\sigma'}d_{A\sigma'}$
is the electronic spin density at the gold site.

The three terms in $\hham_{\rm exch}$ have distinct physical origins.
$J_{el}$ is the direct exchange coupling between the local moment $\Sop_A$
and the electronic spin density $\sop_A$ at the gold site. $J'_p$ is a
spin-dependent charge-redistribution coupling: when the moment is polarised,
the gold on-site energy shifts differently for the two spin species.
$K$ is a hybridisation coupling between the local moment and the
gold-molecule charge transfer. The on-site energies in
Eqs.~\eqref{eq:HAu}-\eqref{eq:eps_A_sigma} and the hybridisation shift in
Eq.~\eqref{eq:selfcons} are the mean-field forms of these three couplings.

The chiral nature of the molecule is encoded in the nearest-neighbour hopping
\begin{align}
H_{n,n+1}
&=
-t_{\rm mol}\left(e^{-\ii\phi/2}d_{n\uparrow}^\dagger d_{n+1,\uparrow}
+
e^{+\ii\phi/2}d_{n\downarrow}^\dagger d_{n+1,\downarrow}
+\text{h.c.}\right)
\nonumber\\
&\quad
+
\ii\lambda_R\sin(\phi/2)\left(d_{n\uparrow}^\dagger d_{n+1,\downarrow}
-
d_{n\downarrow}^\dagger d_{n+1,\uparrow}
+\text{h.c.}\right).
\label{eq:chiral_hop}
\end{align}
The first line is a spin-diagonal phase, pure gauge and physically inert. The
second line is the Rashba spin-flip hopping, which cannot be gauged away and is
the microscopic origin of chirality-dependent spin filtering. The amplitude of
the Rashba term is $\lambda_R\sin(\phi/2)$, where $\phi$ is the twist angle per
bond.

The magnetic moment $\mathbf m = \langle\Sop_A\rangle$ at the gold site is
determined self-consistently from
\begin{equation}
\mathbf m = \frac{1}{2}\tanh\!\left(\frac{\beta|\mathbf B_{\rm
eff}|}{2}\right)\frac{\mathbf B_{\rm eff}}{|\mathbf B_{\rm eff}|},
\label{eq:selfcons}
\end{equation}
with the effective field
\begin{equation}
\mathbf B_{\rm eff}
=
\alpha_A\langle\sop_A\rangle
+
\alpha_1\Delta s\,\hat z
+
\alpha_K\langle\hat{\mathcal N}\rangle\,\hat z
+
b_{\rm static}\hat z,
\label{eq:Beff}
\end{equation}
where $\Delta s = \langle\hat n_{A\uparrow}\rangle-\langle\hat
n_{A\downarrow}\rangle$ is the spin imbalance at the gold site and
$\hat{\mathcal N} = \sum_\sigma(d_{A\sigma}^\dagger d_{B\sigma}+\text{h.c.})$
is the hybridisation operator. We work in the mean-field approximation with
$m_z$ the only non-zero component. This gives the mean-field single-particle
Hamiltonian $\mathbf H_{\rm MF}$ in the site basis $\{|A\uparrow\rangle,
|A\downarrow\rangle, |M_1\uparrow\rangle, \ldots, |M_L\downarrow\rangle\}$.

Here $\beta=1/(k_BT)$ is the inverse thermal energy. The three coefficients
$\alpha_A$, $\alpha_1$, and $\alpha_K$ parametrise the exchange feedback
from the electronic spin density at the gold site, the spin-dependent
charge redistribution, and the hybridisation contribution, respectively.
Their physical origin is the mean-field decoupling of the exchange
interaction in Eq.~\eqref{eq:Hexch}, with $\alpha_A$ corresponding to
$J_{el}$, $\alpha_1$ to $J'_p$, and $\alpha_K$ to $K$; the three coefficients
are kept independent in the numerical work to allow the relative strengths
of the three channels to be varied. The final term is a small static
symmetry-breaking field $b_{\rm static}\hat z$ (with
$b_{\rm static}\approx 5$~meV in the numerical work below) that stabilises
the moment against thermal and quantum fluctuations.

After the mean-field replacement $\Sop_A\to m_z\hat z$, the Hamiltonian
becomes $\mathbf H_{\rm MF}$ in the site basis
$\{|A\uparrow\rangle,\ldots\}$, with the gold on-site energies given by
Eq.~\eqref{eq:eps_A_sigma}, the mean-field molecular on-site energies
\begin{equation}
\varepsilon_{M\sigma} = \varepsilon_{\rm mol} + \sigma\bigl(J'_p+\alpha_{AB}\bigr)m_z,
\label{eq:eps_M_sigma}
\end{equation}
and the hybridisation term $K m_z\sum_\sigma(d_{A\sigma}^\dagger d_{B\sigma}+{\rm h.c.})$.
The parameter $\alpha_{AB}$ entering Eq.~\eqref{eq:eps_M_sigma} is a
phenomenological spin-splitting coefficient for the molecular sites; it
is distinct from $\alpha_A$, $\alpha_1$, $\alpha_K$ above, which act only
at the gold site.

The exact antiunitary symmetry $H^*(\phi)=H(-\phi)$ of the Hamiltonian (see
Ref.~\cite{Alwan2026QSpinterface}) implies the reciprocity relation
Eq.~\eqref{eq:reciprocity_intro} for the retarded Green's function. In the
presence of the leads, the Green's function is itself non-Hermitian, and the
reciprocity breaking that generates the odd part of the spin current is the
same physical effect that underlies the EPs we study below.

\subsection{NEGF and the effective non-Hermitian Hamiltonian}
\label{sec:NEGF_Heff}

The retarded Green's function of the central region is
\begin{equation}
\Gret(\omega) = \left[\omega\mathbf I - \mathbf H_{\rm MF} - \bm\Sigma^R_L(\omega) - \bm\Sigma^R_R(\omega)\right]^{-1}.
\label{eq:Gret}
\end{equation}
In the wide-band limit, each self-energy is
\begin{equation}
\Sigma^R_\alpha(\omega) = -\frac{\ii}{2}\Gamma_\alpha\mathbf P_\alpha,
\label{eq:Sigma_wideband}
\end{equation}
where $\Gamma_\alpha>0$ is the level-width function and $\mathbf P_\alpha$ is the projector onto the sites coupled to lead $\alpha$. The \emph{effective non-Hermitian Hamiltonian} of the junction is
\begin{equation}
H_{\rm eff}(\omega) = \mathbf H_{\rm MF} + \bm\Sigma^R_L(\omega) + \bm\Sigma^R_R(\omega).
\label{eq:Heff}
\end{equation}
Its eigenvalues $\lambda_k$ are the poles of $\Gret(\omega)$ and therefore
control every single-particle transport observable. Since $\mathbf H_{\rm MF}$
is Hermitian and $\bm\Sigma^R_\alpha$ is anti-Hermitian
($(\bm\Sigma^R_\alpha)^\dagger = -\bm\Sigma^R_\alpha$), $H_{\rm eff}$ is
non-Hermitian, with eigenvalues in the complex plane.

The non-Hermiticity is the same source of reciprocity breaking discussed in the
introduction. In the wide-band limit, the anti-Hermitian self-energy breaks the
exact antiunitary symmetry of the closed system, and the resulting
non-reciprocal transport is what produces the odd part of the spin current. The
EP is the singular case where the non-Hermitian structure becomes degenerate.

\subsection{The two EP schemes}
\label{sec:two_schemes}

Consider the projection of $H_{\rm eff}$ onto a two-dimensional subspace spanned by two distinguished states $\{|1\rangle,|2\rangle\}$:
\begin{equation}
H_{\rm eff}^{(2)}
=
\begin{pmatrix}
\tilde\varepsilon_1-\tfrac{\ii}{2}\tilde\gamma_1 & \tilde\alpha\\
\tilde\beta & \tilde\varepsilon_2-\tfrac{\ii}{2}\tilde\gamma_2
\end{pmatrix}.
\label{eq:H2x2}
\end{equation}
The eigenvalues are
\begin{equation}
\lambda_\pm = \bar\varepsilon - \frac{\ii}{2}\bar\gamma \pm \sqrt{\left(\frac{\Delta\tilde\varepsilon}{2} - \frac{\ii}{4}\Delta\tilde\gamma\right)^2 + \tilde\alpha\tilde\beta},
\label{eq:eigenvalues}
\end{equation}
with
\begin{equation}
  \begin{split}
  \bar\varepsilon &= \frac{\tilde\varepsilon_1+\tilde\varepsilon_2}{2},\quad
\Delta\tilde\varepsilon = \tilde\varepsilon_1-\tilde\varepsilon_2,\\
  \bar\gamma &= \frac{\tilde\gamma_1+\tilde\gamma_2}{2},\quad
\Delta\tilde\gamma = \tilde\gamma_1-\tilde\gamma_2.
\end{split}
\end{equation}
At an EP the argument of the square root vanishes. The structure of the
condition depends critically on whether the off-diagonal coupling is Hermitian
or anti-Hermitian.

\paragraph{Scheme~A (Hermitian off-diagonal, $\tilde\beta = \tilde\alpha^*$).}
Then $\tilde\alpha\tilde\beta = |\tilde\alpha|^2>0$, and the discriminant vanishes when
\begin{equation}
\Delta\tilde\varepsilon = 0
\qquad\text{and}\qquad
|\tilde\alpha| = \frac{|\Delta\tilde\gamma|}{4}.
\label{eq:EP_A}
\end{equation}
This is the standard loss-gain EP condition: the diagonal splitting is
cancelled, and the coherent coupling is matched to a quarter of the broadening
difference. It requires \emph{spin-dependent broadening}
($\Delta\tilde\gamma\neq 0$) and a \emph{Hermitian} off-diagonal coupling.

\paragraph{Scheme~B (anti-Hermitian off-diagonal, $\tilde\beta = -\tilde\alpha^*$).}
Then $\tilde\alpha\tilde\beta = -|\tilde\alpha|^2<0$, and the discriminant vanishes when
\begin{equation}
\Delta\tilde\gamma = 0
\qquad\text{and}\qquad
|\tilde\alpha| = \frac{|\Delta\tilde\varepsilon|}{2}.
\label{eq:EP_B}
\end{equation}
This is the complementary condition: the diagonal broadening is symmetric, and
the anti-Hermitian coupling is matched to half the diagonal splitting. It
requires \emph{spin-symmetric broadening} ($\Delta\tilde\gamma=0$) and an
\emph{anti-Hermitian} off-diagonal coupling.

\paragraph{Mixing the schemes: the algebraic obstruction.}
If one naively cancels the diagonal splitting ($\Delta\tilde\varepsilon=0$, the
Scheme~A condition) while using an anti-Hermitian off-diagonal (Scheme~B
structure), then
\begin{equation}
\left(\frac{\Delta\tilde\varepsilon}{2} -
\frac{\ii}{4}\Delta\tilde\gamma\right)^2 - |\tilde\alpha|^2 =
-\left(\frac{\Delta\tilde\gamma}{4}\right)^2 - |\tilde\alpha|^2 < 0,
\label{eq:mixed_impossible}
\end{equation}
which is strictly negative for all real parameters. No EP exists. Both the
diagonal and the off-diagonal must be chosen in the same scheme. Any
configuration that violates this rule gives a strictly negative discriminant and
therefore admits no EP.

\paragraph{Physical interpretation of the two schemes.}
The two schemes are not arbitrary; each is tied to a distinct physical limit of
the junction.

\emph{Scheme~B at the gold interface.} The spin flip at the gold site is
mediated by the electrode: it happens in a virtual tunnelling event, and the
coherence of the process is set by the electronic lifetime in the leads. The
spin-flip coupling therefore appears as an \emph{imaginary} (anti-Hermitian)
off-diagonal element in the effective $2\times 2$. Meanwhile, the two spin
species couple to the leads with equal strength ($\Delta\gamma=0$), because the
leads are non-magnetic and the difference in the couplings of the two spins is
negligible at the gold site. The result is Scheme~B.

\emph{Scheme~A in the molecular interior.} The spin flip on Mol$_1$ is a
coherent on-site process, mediated by the Rashba-active molecular bonds. It
appears as a \emph{real} (Hermitian) off-diagonal element. Meanwhile, the two
spin species experience different broadenings, because the substrate is
spin-polarised (the spinterface mechanism itself provides this spin-dependent
broadening). The result is Scheme~A.

The algebraic obstruction to mixing the schemes reflects a physical fact: you
cannot have an electrode-mediated spin flip (anti-Hermitian) and a
spin-selective substrate (asymmetric broadening) at the same site and expect
the two-state coalescence to occur. Either the spin flip is electrode-mediated
(Scheme~B) or the substrate is spin-selective (Scheme~A), but not both.

The two schemes are mutually exclusive, and this is what makes the gold-EP and
chain-EP regimes genuine classifications rather than merely two points in
parameter space. A configuration that were assignable to neither scheme would
be a numerical coincidence rather than a physical near-EP; a configuration that
were assignable to both would require the diagonal and off-diagonal to be
independently tuned to Scheme~A and Scheme~B respectively, which by
Eq.~\eqref{eq:mixed_impossible} admits no solution. Every near-EP we identify
below is therefore unambiguously of one type or the other, and this is a
stronger statement than the residual-gap scaling alone can provide.

\paragraph{Visual summary of the two EP schemes.}
Figure~\ref{fig:EPschematic} provides a visual summary of the model and
of the two EP schemes derived above.
The figure shows the junction geometry with the two EP mechanisms
highlighted. The red star at the gold site marks Scheme~B: the spin flip
at Au is electrode-mediated, hence anti-Hermitian, and it requires the
leads to be spin-symmetric ($\Delta\gamma_{\rm gold}=0$). The green
star at the first molecular site marks Scheme~A: the spin flip on
Mol$_1$ is a coherent on-site Rashba process, hence Hermitian, and it
requires a spin-dependent substrate broadening
($\Delta\gamma_{\rm chain}\neq0$), shown by the green bracket over the
helix. The two coloured EP stars are connected by thin pointer lines to
the corresponding callout boxes above the device, and each callout
quotes the associated EP condition. The dark-red box states the
algebraic obstruction of Eq.~\eqref{eq:mixed_impossible}: because
$-(\Delta\gamma/4)^2-|\alpha|^2<0$ for all real parameters, the two
schemes cannot be mixed and no EP exists on any parameter path that
combines the diagonal cancellation of one scheme with the off-diagonal
structure of the other. 
%
The two lower rows below the device display the self-consistent spin
texture $\langle\mathbf{s}_n\rangle$ at the reference parameters.
Their content is discussed in Sec.~\ref{sec:gold}; here we note only
that the strip is drawn from the same NEGF calculation that is used
throughout the paper, and its presence in the schematic is meant to
remind the reader that the EP physics is embedded in a well-defined,
quantitative transport problem, not an isolated algebraic construction.

\subsection{Petermann factor and its divergence}
\label{sec:petermann_math}

For a non-Hermitian matrix $H_{\rm eff}$, the right and left eigenproblems are
\begin{equation}
H_{\rm eff}|R_k\rangle = \lambda_k|R_k\rangle,\qquad
\langle L_k|H_{\rm eff} = \lambda_k\langle L_k|,
\label{eq:left_right}
\end{equation}
with biorthogonal eigenvectors
$\langle L_j|R_k\rangle=\delta_{jk}\langle L_k|R_k\rangle$. The Petermann
factor of eigenstate $k$ is
\begin{equation}
K_k = \frac{\langle L_k|L_k\rangle\langle R_k|R_k\rangle}{|\langle L_k|R_k\rangle|^2},
\label{eq:petermann}
\end{equation}
so that $K_k=1$ in the Hermitian limit, where $\hat L_k=\hat R_k$ after
normalisation.

Near an EP, both the biorthogonal overlap and the eigenvalue gap vanish. For a
generic $2\times2$ non-Hermitian matrix with individually normalised left and
right eigenvectors, the overlap scales \emph{linearly} in the eigenvalue gap,
\begin{equation}
|\langle \hat L_k|\hat R_k\rangle| \propto \delta,
\label{eq:overlap_scaling}
\end{equation}
where $\delta=|\lambda_1-\lambda_2|$ is the gap. The simplest example is
$H=\begin{pmatrix}0&1\\\varepsilon&0\end{pmatrix}$, for which
$\delta=2\sqrt{\varepsilon}$, $|R\rangle\propto(1,\sqrt{\varepsilon})$,
$\langle L|\propto(\sqrt{\varepsilon},1)$, and, after individually
normalising both vectors to unit $\ell^2$ norm,
$|\langle\hat L|\hat R\rangle|=2\sqrt{\varepsilon}/(1+\varepsilon)
=\delta/(1+\varepsilon)\simeq\delta$ for small $\varepsilon$. Substituting
into Eq.~\eqref{eq:petermann},
\begin{equation}
K_k \propto \frac{1}{|\langle \hat L_k|\hat R_k\rangle|^2}
\propto \frac{1}{\delta^2}.
\label{eq:K_divergence}
\end{equation}
This is the correct scaling for the individually normalised convention used
throughout this paper; the alternative $K\propto1/\delta$ found in parts of the
literature corresponds to a biorthogonal-normalisation convention, which we do
not use.

In the full $N\times N$ junction the effective $2\times2$ is dressed by virtual
transitions to the rest of the spectrum, which renormalises both the effective
splitting and the residual gap. Empirically, the peak Petermann factor at the
dressed near-EP is well described by
\begin{equation}
K_{\rm max} \sim \frac{\Delta\varepsilon_{\rm eff}^2}{\delta_{\rm res}^2},
\label{eq:K_max_empirical}
\end{equation}
where $\Delta\varepsilon_{\rm eff}$ is the effective diagonal splitting of the
dressed pair and $\delta_{\rm res}$ its residual gap. In the isolated limit
$\Delta\varepsilon_{\rm eff}\to\Delta\varepsilon_{\rm bare}$ and the empirical
peak reduces to $K_{\rm max}\to(\Delta\varepsilon_{\rm bare}/\delta_{\rm
res})^2$, which is the bare two-state result. The ratio
$\Delta\varepsilon_{\rm eff}^2/\Delta\varepsilon_{\rm bare}^2$ is therefore
the suppression factor by which the dressing reduces $K_{\rm max}$ relative to
the undressed limit.
The empirical scaling is a direct consequence of the second-order Schur
complement: the bath corrections shift the diagonal splitting to $\Delta
\varepsilon_{eff}$ and the residual gap to $\delta_{res} \approx
|H_{PQ}|^{2}/|\bar{\varepsilon_{0}} - \varepsilon_{bath}|$, and substituting
both into the two-state Petermann formula gives~\eqref{eq:K_max_empirical} up
to a factor of order unity.

\paragraph{Numerical evaluation.}
A numerical subtlety worth highlighting is to use the standard
biorthogonally-normalised left eigenvectors $\mathbf V_L=\mathbf
V_R^{-1\dagger}$, which give $\langle L_k|R_k\rangle=1$ by construction and
therefore $K_k=1$ everywhere. The correct algorithm extracts the left
eigenvectors independently via the nullspace of $(H_{\rm
eff}^\dagger-\lambda_k^*\mathbf I)$, normalises them to unit $\ell^2$ norm, and
evaluates $K_k$ from Eq.~\eqref{eq:petermann}.

\subsection{Parameter regime and numerical implementation}
\label{sec:parameters}

All numerical results use the following parameters, unless otherwise noted:
$t_{AB}=t_{\rm mol}=0.5$~eV, $J_{\rm el}=-0.5$~eV, $J'_p=-0.3$~eV,
$\alpha_{AB}=0.5$~eV, $\Gamma_0=0.5$~eV, $\Delta\gamma=0.4$~eV, $T=300$~K,
$\phi=60^\circ$.

A note on the choice of $\Gamma_0$. The gold-EP figures (Figs.~\ref{fig:EP_IV}
and~ \ref{fig:chirality_crossover}) use a gold-site broadening
$\Gamma_0=0.5$~eV. This larger broadening places the near-EP on a visible
transport resonance. The chain-EP figures (Figs.~\ref{fig:C1}-\ref{fig:C3})
retain the baseline gold-site coupling $\Gamma_L=0.1$~eV of the underlying
model, with the chain broadenings $\gamma_{\rm
chain,\uparrow/down}=\Gamma_0^{\rm chain}\pm\Delta\gamma/2$ using
$\Gamma_0^{\rm chain}=0.5$~eV. The two figure families therefore probe
different aspects of the same near-EP mechanism: the gold figures emphasise the
transport signature at a fixed bias, while the chain figures preserve the
collective-scaling structure that requires weak leads and is more sensitive to
$L$. The choice is a deliberate operating-point decision, not an inconsistency
in the underlying model.

For the EP figures:
\begin{itemize}
\item \textbf{Gold-level near-EP}: $\varepsilon_A=1.5$~eV, $t_{AB}=0.5$~eV,
  $L=6$, $\Delta\gamma=0.4$~eV.
\item \textbf{Chain-level near-EP}: $\varepsilon_A=1.5$~eV, $t_{AB}=0.5$~eV,
  $L=6$ reference (unless explicitly varied), $\lambda_R^{\rm chain}=0.05$~eV
  (weakened from the physical value $0.6$~eV), $\theta_R=\phi$ (helical limit).
\end{itemize}
The weakened Rashba amplitude for the chain-level study is discussed in more
detail in Sec.~\ref{sec:chain} and in the Conclusion; it is a controlled
parameter for isolating the collective mechanism, not a prediction for the
specific molecule.
At each parameter point we compute $\Gret(\omega)$ on a $2000$-point frequency
grid, propagate to $\Gless(\omega)$ via the Keldysh equation, and compute the
transport observables via the Landauer formula. The eigenvalue tracking uses
the left/right eigenvectors, with the Petermann factor computed by the
nullspace algorithm described above.

The self-consistent moment $m_z$ is computed at $V=1$~V without any EP
modification, giving $m_z\approx 0.438$, which sets the bare gold splitting to
$|\Delta\varepsilon_A|\approx 0.482$~eV.

\section{Gold-Level Near-Exceptional Point}
\label{sec:gold}

We first study the near-exceptional point that emerges at the
electrode-molecule interface when the two gold spin states coalesce. This
regime is a controlled two-state problem in which the EP mechanism can be
analysed cleanly, and it is the natural starting point for the analysis
because the two-state approximation is best justified here.

The reader should note from the outset that $\varepsilon_A = 1.5$~eV sits
roughly $1$~eV above the bias-window edge at the operating bias $V\lesssim
1$~V. The gold resonance is therefore \emph{outside} the window, and the EP
modifies only the exponential tails of the transmission rather than a resonant
peak. This is why the transport I-V curves (Fig.~\ref{fig:EP_IV}) show a
modest fractional change across the EP, even though the spectral signatures
(Fig.~\ref{fig:G1} and Fig.~\ref{fig:G2}) are sharp. The spectral signatures
are the primary evidence for the EP; the transport signature is a secondary
corroboration whose magnitude is set by the operating point. The choice
$\varepsilon_A=1.5$~eV is deliberate: it isolates the gold doublet from the
molecular band and reduces the dressing-induced residual gap, but the price is
that the transport signature is weak. A different choice of $\varepsilon_A$
nearer to the Fermi level would strengthen the transport signature at the cost
of a larger residual gap. This trade-off is quantified in
Sec.~\ref{sec:gold_petermann}.

Section~\ref{sec:gold_setup} states the two-state projection and the bare EP
condition. Section~\ref{sec:gold_tracking} introduces the gold-character
tracking method and presents the resulting eigenvalue trajectories
(Fig.~\ref{fig:G1}). Section~\ref{sec:gold_petermann} quantifies the Petermann
enhancement at the near-EP (Fig.~\ref{fig:G2}) and the second-order dressing
analysis that explains the residual gap. The transport response at fixed bias
and the chirality-selective crossover are deferred to
Secs.~\ref{sec:gold_IV} and~\ref{sec:chirality_crossover_sec} below.
%

\subsection{Setup and EP condition}
\label{sec:gold_setup}

The relevant subspace is $\{A\uparrow,A\downarrow\}$. The spin splitting is
\begin{equation}
\Delta\varepsilon_A = |(J_{\rm el}+2J'_p)m_z|\approx 0.482~\text{eV}.
\label{eq:Delta_A}
\end{equation}
The spin-flip process is mediated by an anti-Hermitian self-energy at the gold
site,
\begin{equation}
\Sigma_{\rm sf}^R = -\frac{\ii}{2}g_{\rm sf}e^{\ii\varphi}
\bigl(d_{A\uparrow}^\dagger d_{A\downarrow}+\text{h.c.}\bigr),
\label{eq:Sigma_sf_gold}
\end{equation}
which contributes an off-diagonal $|\tilde\alpha| = g_{\rm sf}/2$ in the
effective $2\times 2$. The leads are spin-symmetric
($\gamma_{L\uparrow}=\gamma_{L\downarrow}=\Gamma_0$), so $\Delta\gamma=0$. The
gold doublet therefore realises Scheme~B, and the predicted EP is at
\begin{equation}
g_{\rm sf}^{\rm EP, bare} = |\Delta\varepsilon_A|\approx 0.48~\text{eV}.
\label{eq:gold_EP_prediction}
\end{equation}

\paragraph{Physical origin of the anti-Hermitian spin flip.}
The self-energy Eq.~\eqref{eq:Sigma_sf_gold} is not an arbitrary choice; it is
the effective description of a spin-flip process in which an electron from the
gold site hops into the electrode, flips its spin (due to the strong Rashba
spin-orbit coupling at the Au surface), and hops back. Because the intermediate
state is a virtual state in the continuum of the electrode, the process
acquires an imaginary character: the effective spin-flip coupling is
\emph{anti-Hermitian}. This is a well-known result in the NEGF theory of
magnetic impurities in metals, and it is the natural physical mechanism at the
spinterface. The magnitude $g_{\rm sf}$ is proportional to the electrode's
spin-orbit coupling and to the local density of states at the gold site.

\subsection{Eigenpair tracking}
\label{sec:gold_tracking}

Because the gold doublet is dressed by the molecular chain, the two gold-like
eigenvalues are not necessarily the two lowest-energy or two highest-energy
states of $H_{\rm eff}$. We identify them by their \emph{gold character}
\begin{equation}
  \begin{split}
  S_k &= \text{Re}\left[\langle L_k|\hat P_A|R_k\rangle\right],\\
  \hat P_A &= |A\uparrow\rangle\langle A\uparrow|+|A\downarrow\rangle\langle
A\downarrow|,
\label{eq:gold_character}
\end{split}
\end{equation}
which measures the fraction of eigenvector $k$ localised on the gold site. The
two eigenvalues with the largest $|S_k|$ form the gold-like pair. This tracking
method is robust: the gold character varies smoothly with $g_{\rm sf}$ and
vanishes only when the eigenstate is entirely delocalised onto the molecule,
which does not occur in our parameter range.

\begin{figure}[htbp]
\centering
\includegraphics[width=0.9\linewidth]{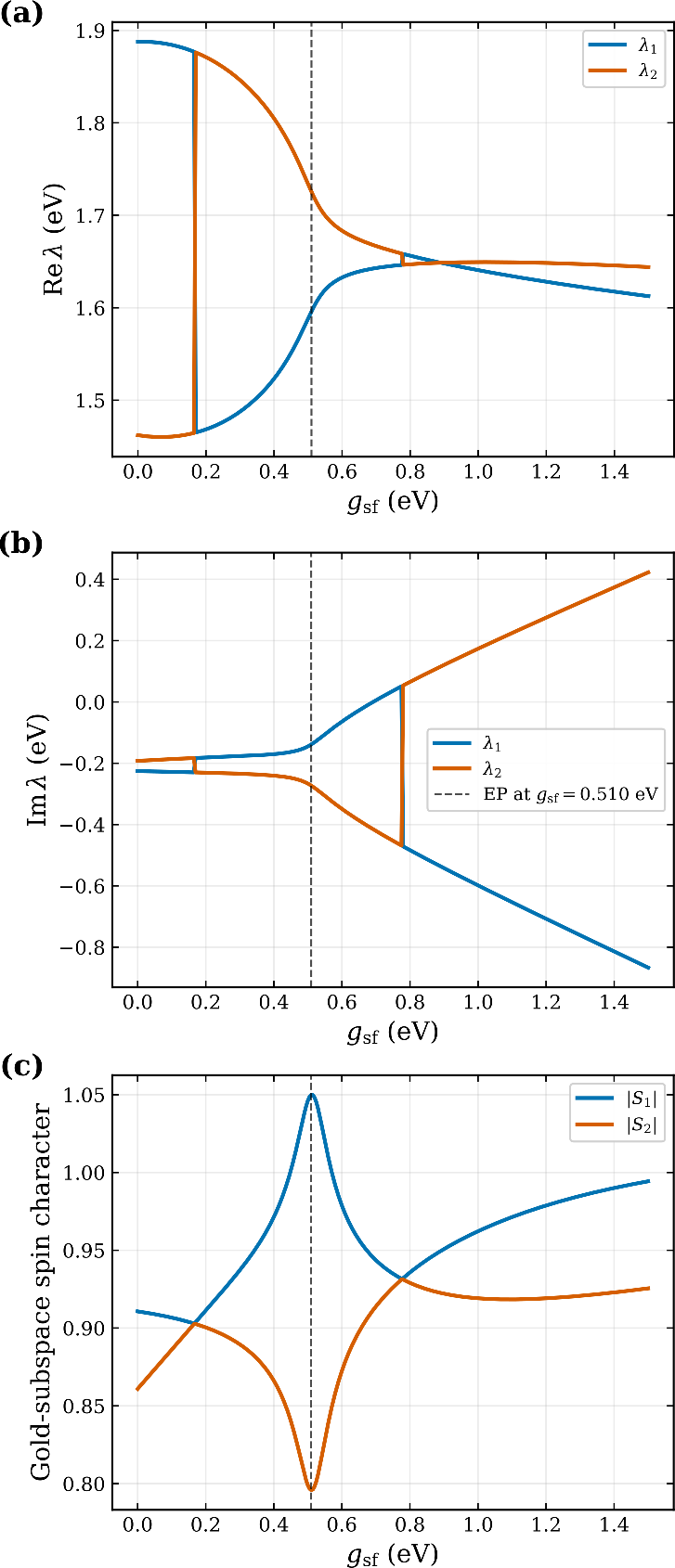}
\caption{Gold-level near-EP eigenpair tracking. (a) Real parts of the two
gold-like eigenvalues as a function of the spin-flip coupling $g_{\rm sf}$. The
two branches approach each other, reach a minimum separation at the near-EP
(vertical dashed line at $g_{\rm sf}\approx 0.51$~eV), and split again. The
minimum separation $\approx 0.19$~eV is a consequence of the dressing of the
gold doublet by the molecular chain. (b) Imaginary parts. Below the near-EP the
two eigenvalues have distinct imaginary parts (differential lifetimes of the
two spin channels); at the near-EP they merge; above they split again with an
exchanged sign structure, the sign reversal occurring at $g_{\rm sf}\approx
0.7$-$0.8$~eV, in the broken phase just above the near-EP. (c) Biorthogonal
gold character $|S_k|$ of the two tracked eigenvectors. Far from the near-EP,
the two eigenvectors are nearly pure $A\uparrow$ and $A\downarrow$. At the
near-EP they mix and become equal-weight superpositions. Parameters: $L=6$,
$\Delta\gamma=0.4$~eV, $\varepsilon_A=1.5$~eV, $V=1$~V.}
\label{fig:G1}
\end{figure}

Figure~\ref{fig:G1} shows three complementary views of the tracked pair as
$g_{\rm sf}$ is varied. Panel~(a) shows the real parts: the two eigenvalues
approach each other, reach a minimum separation at $g_{\rm sf}\approx
0.51$~eV, then split again. The minimum separation is $\approx 0.19$~eV, not
zero, but a residual that reflects the dressing of the gold doublet by the
molecular chain (Sec.~\ref{sec:gold_petermann}). The shift of the observed EP
from the bare prediction ($0.510$~eV vs.\ $0.482$~eV) is small: a dressing
factor of $g_{\rm sf}^{\rm EP}/|\Delta\varepsilon_A|=1.06$, i.e.\ a
$\approx 5.9\%$ shift, also attributable to dressing.

Panel~(b) shows the imaginary parts of the two eigenvalues. Below the near-EP,
the two eigenvalues have distinct imaginary parts: the exchange field
polarises the two spin channels differently, so their lifetimes (inverse
imaginary parts) differ. At the near-EP they merge; above they split again
with an exchanged sign structure. The exchange of imaginary parts is a
hallmark of the EP: the two eigenvectors, which were localised on $A\uparrow$
and $A\downarrow$ below the near-EP, coalesce into a single superposition at
the near-EP, and then re-separate into new combinations above. Numerically,
the sign reversal occurs at $g_{\rm sf}\approx 0.7$-$0.8$~eV, i.e.\ in the
broken phase just above the near-EP rather than at the near-EP itself. 

Panel~(c) shows the biorthogonal gold character $|S_k|$ of the two tracked
eigenvectors. Far from the near-EP, one eigenvector is predominantly
$A\uparrow$ and the other $A\downarrow$, so each has large gold character. As
the near-EP is approached, the two eigenvectors mix; at the near-EP they
coalesce into a single vector with equal weight on both spin states. The
collapse of the gold character difference is the eigenvector signature of the
EP.

\subsection{Petermann enhancement and the dressing of the near-EP}
\label{sec:gold_petermann}

\begin{figure}[htbp]
\centering
\includegraphics[width=1.0\linewidth]{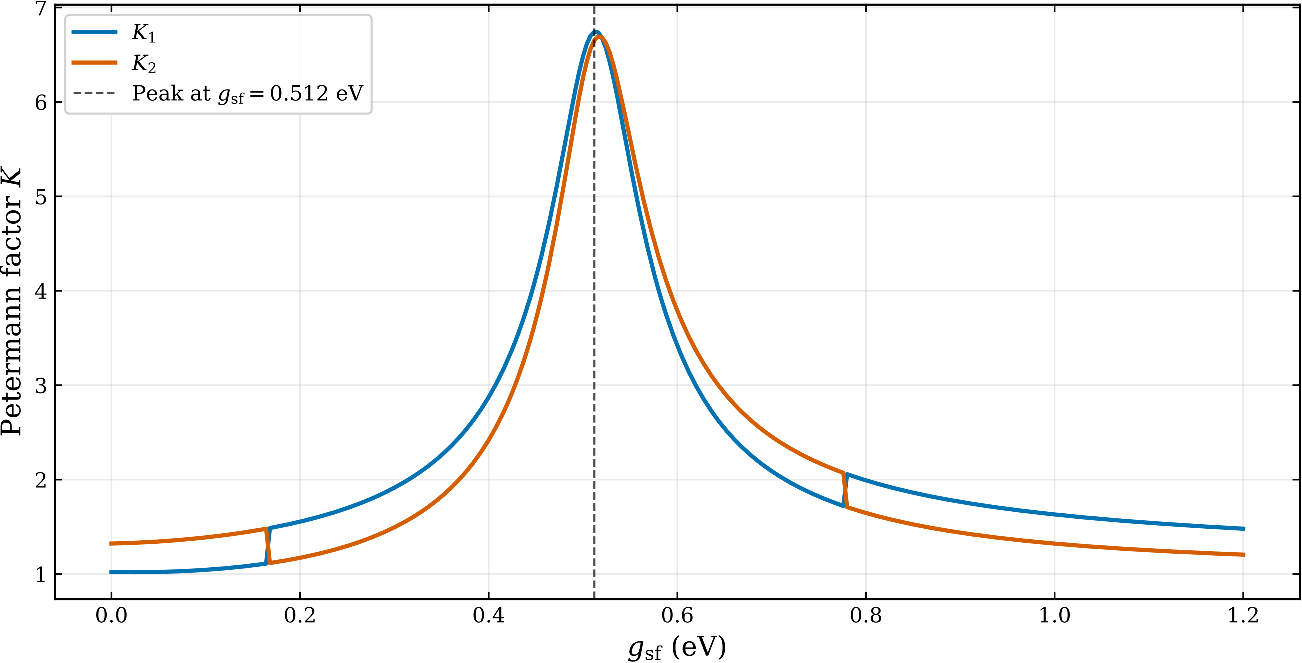}
\caption{Gold-level near-EP Petermann factor. Both $K_1$ and $K_2$ rise
sharply at the near-EP (vertical dashed line) and fall again outside. The peak
value is $K\approx 6.7$, which is consistent with the near-EP residual gap of
$\sim 0.19$~eV: a strict EP would have $K\to\infty$, but the dressing of the
gold doublet by the molecular chain prevents the last bit of coalescence. The
divergence is symmetric in $g_{\rm sf}-g_{\rm sf}^{\rm EP}$, as expected from
$K_{\rm max}\sim\Delta\varepsilon_{\rm eff}^2/\delta_{\rm res}^2$. Parameters
as in Fig.~\ref{fig:G1}.}
\label{fig:G2}
\end{figure}

The Petermann factor is the most sensitive diagnostic of the near-EP.
Figure~\ref{fig:G2} shows $K_1$ and $K_2$ for the two tracked eigenstates. Far
from the near-EP, $K\approx 1$, indicating Hermitian behaviour. As the near-EP
is approached, $K$ rises sharply to a peak of $K\approx 6.7$ at
$g_{\rm sf}\approx 0.51$~eV, then falls again. The width of the peak is set by
the residual gap: $K$ is enhanced only when
$|\lambda_1-\lambda_2|\lesssim\delta_{\rm res}$, which for the gold parameters
is a half-width at half-maximum of $\sim 0.08$~eV (full width $\sim 0.16$~eV)
around the near-EP.

\paragraph{Dressing of the near-EP.}
The gold doublet is not perfectly isolated from the rest of the spectrum. Even
at $\varepsilon_A=1.5$~eV, $\approx 1.5$~eV above the molecular band edge, the
direct Au-Mol$_1$ hopping $t_{AB}=0.5$~eV generates virtual transitions that
shift and broaden the gold levels. From second-order perturbation theory,
\begin{equation}
\delta_{\rm res}^{\rm gold}\sim
\frac{t_{AB}^2}{|\varepsilon_A-\varepsilon_{\rm mol}|}\sim 0.25~\text{eV},
\label{eq:residual_gold}
\end{equation}
which is of the same order  of the observed $0.186$~eV. The expected accuracy
of a second-order estimate when the perturbing hopping $t_{AB}$ is not small
compared to the detuning. In the isolated-gold limit $t_{AB}\to 0$, the
residual would vanish and the near-EP would become a strict EP. The residual
gap is therefore a controlled diagnostic of the isolation of the gold doublet.

The Petermann peak value $K_{\rm max}=6.7$ is consistent with the residual-gap
analysis of Sec.~\ref{sec:petermann_math}: from Eq.~\eqref{eq:K_max_empirical}
with $\Delta\varepsilon_{\rm eff}\approx\Delta\varepsilon_A\approx 0.4817$~eV
at the near-EP,
\begin{equation}
K_{\rm max}\approx \frac{\Delta\varepsilon_A^2}{\delta_{\rm res}^2}\approx
\frac{0.4817^2}{0.1864^2}\approx 6.7,
\label{eq:K_gold_estimate}
\end{equation}
in excellent agreement with the observed $6.73$. This quantitative match
confirms that the near-EP is a genuine non-Hermitian feature, and that the
residual gap is its dominant limiter. It is also a useful consistency check: if
the residual were dominated by a different effect (e.g., an external magnetic
field, or phonon coupling), the observed peak $K$ would not match the
empirical scaling.

\subsection{Transport I-V characteristics at the near-EP}
\label{sec:gold_IV}

\begin{figure}[htbp]
\centering
\includegraphics[width=0.8\linewidth]{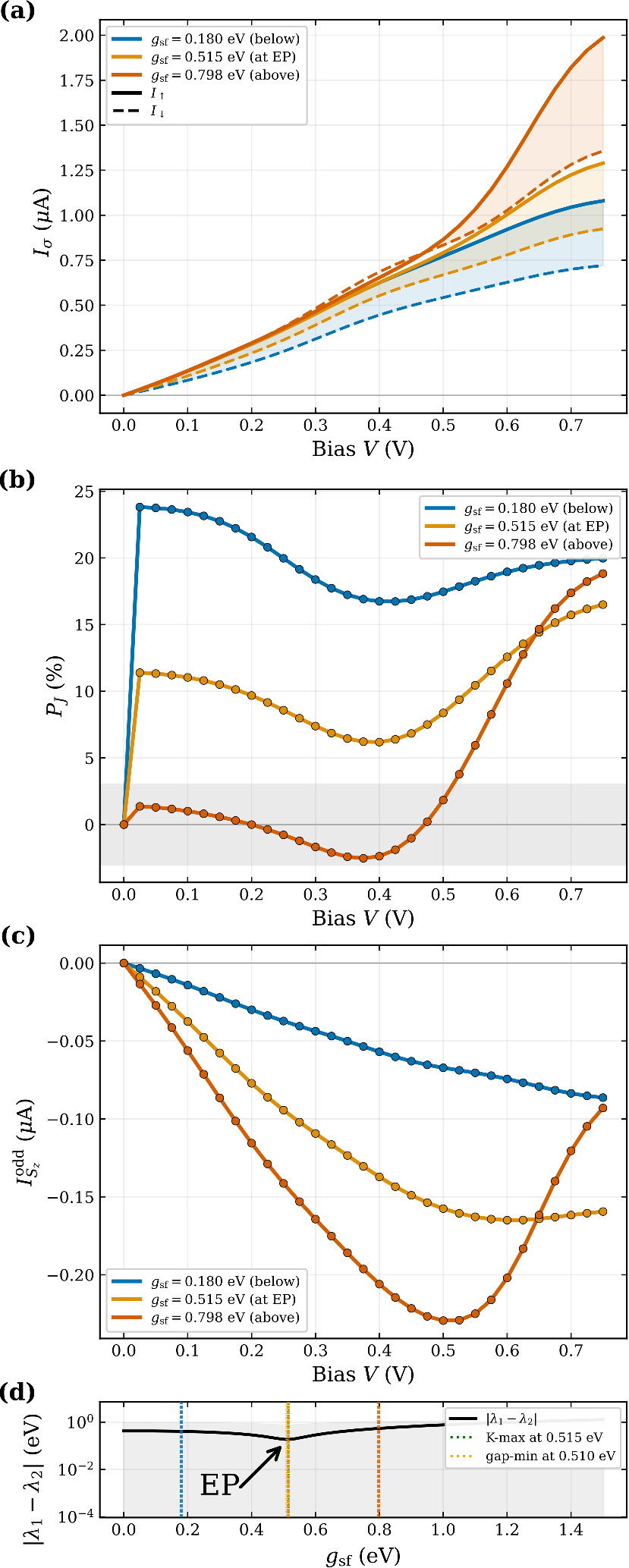}
\caption{Exceptional-point modification of the CISS I-V characteristics. (a)
Spin-resolved currents $I_\uparrow$ (solid) and $I_\downarrow$ (dashed), in
$\mu$A, vs.\ bias $V$ at three couplings: below the near-EP
($g_{\rm sf}=0.18$~eV, blue), at the near-EP ($g_{\rm sf}=0.51$~eV, orange),
and above ($g_{\rm sf}=0.79$~eV, red). At the near-EP the two spin channels
nearly coalesce. (b) Current polarisation $P_J(V)$ for the same three
couplings. Below the near-EP, $P_J$ is large and positive across the bias
window; at the near-EP, it is suppressed to $\sim 8\%$ in the low-bias regime
($V\lesssim0.5$~V) with weaker suppression at higher bias; above the near-EP,
it recovers at higher bias ($V\gtrsim0.7$~V) but is strongly suppressed at
low bias, becoming slightly negative at $V=0.25$~V. The sign change that
accompanies the achiral-to-chiral crossover. The shaded horizontal band marks
the achiral-crossing region $|P_J|<3\%$. (c) Odd (chiral) part of the spin
current, $I_{S_z}^{\rm odd}(V)$, which grows monotonically with $g_{\rm sf}$
across the whole bias window. (d) Gold-like eigenpair gap vs.\ $g_{\rm sf}$,
with the three couplings marked by vertical dashed lines. The gap minimum
(residual $\approx 0.19$~eV) coincides with the near-EP. Parameters: $L=6$,
$\Delta\gamma=0.4$~eV, $\varepsilon_A=1.5$~eV, $\phi=60^\circ$, $T=300$~K.}
\label{fig:EP_IV}
\end{figure}

Figure~\ref{fig:EP_IV} shows how the near-EP modifies the transport I-V
characteristics. Panel~(a) shows the spin-resolved currents $I_\uparrow$ and
$I_\downarrow$ (in $\mu$A) at three couplings: below the near-EP
($g_{\rm sf}=0.18$~eV), at the near-EP ($g_{\rm sf}=0.51$~eV), and above the
near-EP ($g_{\rm sf}=0.79$~eV). Below the near-EP, the solid and dashed curves
for $I_\uparrow$ and $I_\downarrow$ are visibly separate, indicating
spin-resolved transport. At the near-EP, the two curves sit nearly on top of
each other, indicating the coalescence of the two spin channels. Above the
near-EP, they split again, but with a different sign structure.

Panel~(b) shows the current polarisation $P_J(V)$ for the three couplings.
Below the near-EP, $P_J$ is large and positive across the bias window, peaking
at $\sim 23\%$ at $V=0.1$~V. At the near-EP, $P_J$ is reduced to $\sim 8\%$
in the low-bias regime ($V \lesssim 0.5$~V), with the suppression becoming
weaker as the bias window widens toward the gold resonance. Above the
near-EP, $P_J$ recovers at higher bias ($V \gtrsim 0.7$~V) but is strongly
suppressed at low bias, becoming slightly negative at $V=0.25$~V. The
signature of the chirality-dominated regime first overtaking the achiral
spin-valve contribution (see Fig.~\ref{fig:chirality_crossover}).

Panel~(c) shows the odd (chiral) part of the spin current,
$I_{S_z}^{\rm odd}(V)$, decomposed using the even/odd scheme. The
odd part grows monotonically with $g_{\rm sf}$ across the whole bias window.
Panel~(d) shows the gold-like eigenpair gap as a function of $g_{\rm sf}$,
with the three couplings marked by vertical dashed lines. The gap minimum
coincides with the near-EP and with the peak of $P_J$ suppression in
panel~(b).


\subsection{Chirality-selective transport and the three crossover points}
\label{sec:chirality_crossover_sec}

The most striking transport result is that the anti-Hermitian spin-flip
coupling is \emph{chirality-selective}: it coherently mixes the two gold spin
states, converting achiral spin-valve transport into chiral CISS transport.
This is the dynamical analogue of the even/odd decomposition of the spin
current established in Ref.~\cite{Alwan2026QSpinterface}: the spin current is
split into a part that is even under chirality reversal (the achiral spin-valve
baseline) and a part that is odd (the genuine CISS signal). Here, however, the
decomposition is not just a bookkeeping tool; it reflects the physical
conversion of achiral to chiral response as the spin-flip coupling increases.

To quantify the conversion, we decompose the spin current into even and odd
parts with respect to chirality reversal $\phi\to-\phi$:
\begin{align}
I_{S_z}^{\rm even}(\phi) &=
\tfrac12\left[I_{S_z}(+\phi)+I_{S_z}(-\phi)\right],\label{eq:even_def}\\
I_{S_z}^{\rm odd}(\phi) &=
\tfrac12\left[I_{S_z}(+\phi)-I_{S_z}(-\phi)\right].\label{eq:odd_def}
\end{align}
The even part is invariant under $\phi\to-\phi$ and originates from the achiral
spin-valve effect of the gold moment. The odd part changes sign under
$\phi\to-\phi$ and is the true CISS signal. The decomposition is exact:
$I_{S_z}=I_{S_z}^{\rm even}+I_{S_z}^{\rm odd}$.

\begin{figure}[htbp]
\centering
\includegraphics[width=\linewidth]{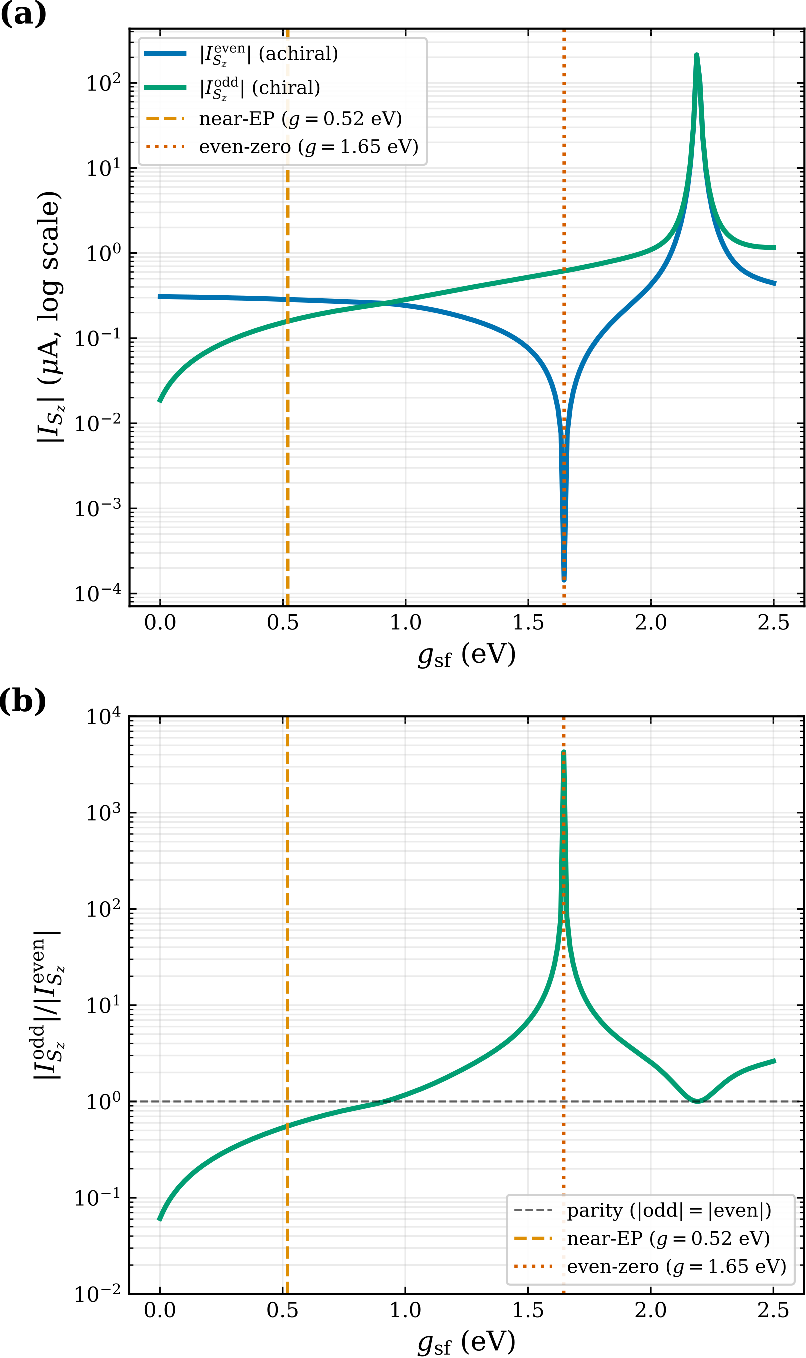}
\caption{Chirality crossover at the gold-level near-EP. (a) Absolute values of
  the achiral even part $|I_{S_z}^{\rm even}|$ and the chiral odd part
  $|I_{S_z}^{\rm odd}|$ of the spin current, in $\mu$A, vs.\ $g_{\rm sf}$ at
  $V=0.5$~V (logarithmic vertical scale). The near-EP ($g=0.51$~eV) is marked
  by an orange dashed line; the achiral extinction point ($g=1.63$~eV) by a red
  dotted line. The achiral part drops by more than two orders of magnitude at
  the extinction point. (b) Chirality selectivity ratio $|I_{S_z}^{\rm
  odd}|/|I_{S_z}^{\rm even}|$ on a logarithmic scale. The ratio crosses unity
  at the first parity point ($g\approx 0.86$~eV), is driven to a formal
  divergence at the achiral-extinction point ($g=1.63$~eV), and returns to
  order-one values through a second parity crossing at $g\approx 2.2$~eV. The
three crossover points, i.e., the near-EP onset, first parity, and extinction,
are the essential experimental signatures of the chirality-selective
mechanism.}
\label{fig:chirality_crossover}
\end{figure}

Figure~\ref{fig:chirality_crossover} shows the two components and their ratio
as $g_{\rm sf}$ varies at $V=0.5$~V. The ratio grows monotonically from
$0.062$ at $g_{\rm sf}=0$ through a first parity crossing near the near-EP,
diverges at the achiral-extinction point, and returns to order-one values
through a second parity crossing above the extinction point.


Three crossover points emerge from this structure.

\paragraph{(i) Onset of chirality dominance at the near-EP.}
At $g_{\rm sf}\approx 0.51$~eV, the ratio $|{\rm odd}|/|{\rm even}|$ rises to
$0.56$ (from $0.062$ at $g=0$), and the total polarisation $P_J$ drops from
$22.3\%$ to $8.5\%$. This is the transition from achiral-dominated to
chirality-relevant transport. The near-EP location and the onset of chirality
dominance coincide by construction of the mechanism: the spin-flip coupling
both drives the near-EP and converts achiral into chiral character.

\paragraph{(ii) First parity crossing.}
At $g_{\rm sf}\approx 0.86$~eV, the two contributions are equal
($|{\rm odd}|=|{\rm even}|$, ratio $=1$), marking the first parity crossing.
Above this coupling, the chiral part dominates and the total polarisation
crosses zero and continues to decrease monotonically; at $g_{\rm sf}=1.00$~eV
the polarisation is $-2.56\%$, and it drops further to $-19.7\%$ at
$g_{\rm sf}=1.63$~eV. 
A second parity crossing occurs at $g_{\rm sf}\approx 2.19$~eV, after the
achiral-extinction spike has pushed the ratio through a local minimum. Neither
crossing is an EP; both are symmetry points where the two contributions
balance.

\paragraph{(iii) Complete achiral extinction.}
At $g_{\rm sf}\approx 1.63$~eV, the achiral spin-valve signal vanishes (its
absolute value drops by a factor of $110$ from its value at $g=0$), while the
chiral odd part is at its strongest. The ratio $|{\rm odd}|/|{\rm even}|$
formally diverges here. This is the completion of the chirality crossover: at
$g_{\rm sf}=1.63$~eV the junction is purely chiral in its spin-transport
response. Of the three crossover points, the extinction point is the most
quantitatively robust: its location is confirmed to four significant figures
by the full $N\times N$ diagnostic as the minimum of $|I_{S_z}^{\rm even}|$
over a fine scan, and it is therefore the most testable experimental
prediction of the chirality-selective mechanism.

The ratio of the extinction point to the near-EP is $1.63/0.51\approx 3.2$.
This dimensionless factor is a characteristic of the spinterface mechanism,
determined by the interplay between the achiral spin-valve effect of the gold
moment and the Rashba-driven chirality of the chain.

\paragraph{Analytic model of the crossover.}
In the effective two-state model the achiral spin current is proportional to
the diagonal splitting $\Delta\varepsilon$ and the chiral spin current to the
off-diagonal coupling $g_{\rm sf}$. Setting $g\equiv g_{\rm sf}$ and
$D\equiv\Delta\varepsilon$, the two currents scale as
\begin{equation}
I_{S_z}^{\rm even}\propto \frac{D}{\sqrt{D^2+g^2}},\qquad
I_{S_z}^{\rm odd}\propto \frac{g}{\sqrt{D^2+g^2}},
\label{eq:two_state_chirality}
\end{equation}
so the ratio is
\begin{equation}
\frac{|I_{S_z}^{\rm odd}|}{|I_{S_z}^{\rm even}|} = \frac{g}{D}.
\label{eq:chirality_ratio}
\end{equation}
At $g=D$ the ratio equals $1$, this is the first parity crossing, which coincides with
the near-EP in the two-state model. At $g=2D$ the ratio is $2$, and at $g=3D$
it is $3$: the two-state model has no extinction point and no second parity
crossing, because it contains only the achiral and chiral contributions and no
chain dressing. The extinction point observed at $g_{\rm sf}\approx 1.63$~eV
and the second parity crossing at $g_{\rm sf}\approx 2.19$~eV are therefore
\emph{beyond} the two-state model. They arise because the chain dressing
reduces the effective splitting $\Delta\varepsilon_{\rm eff}$ below the bare
$D$, so that the ratio $g/D_{\rm eff}$ diverges when
$\Delta\varepsilon_{\rm eff}\to 0$. The two-state model thus captures the

ratio beyond it, but the specific extinction location requires the full
$N\times N$ treatment. The observed extinction-to-parity ratio
$1.63/0.86\approx 1.9$ is a quantitative measure of the chain-dressing
amplification of the two-state prediction.

\subsection{Polarisation and transmission collapse}
\label{sec:gold_collapse}

\begin{figure}[htbp]
\centering
\includegraphics[width=0.8\linewidth]{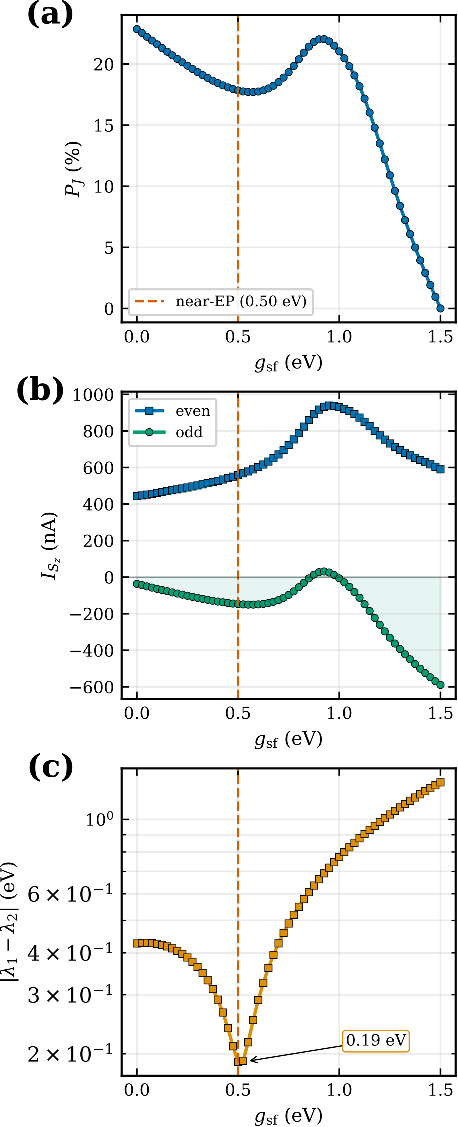}
\caption{Gold-level near-EP: polarisation collapse, spin-current
decomposition, and coalescence diagnostic ($L=6$,
$\Delta\gamma=0.4$~eV, $V=1$~V). (a) Current polarisation $P_J$
vs.\ $g_{\rm sf}$. The polarisation is maximal at $g=0$ (achiral
spin-valve), decreases through the near-EP (vertical dashed line,
$g_{\rm sf}\approx 0.51$~eV), reaches its lowest values in the
parity region ($g_{\rm sf}\approx 0.9$--$1.0$~eV), and flattens out
at larger $g_{\rm sf}$ where the chiral part dominates.
(b) Even (achiral, blue dashed) and odd (chiral, green solid,
shaded) parts of the spin current on a common axis, in nA. The
even part is largest at $g=0$ and decreases monotonically; the odd
part starts near zero and grows monotonically. The two curves cross
at the parity point where $|I_{S_z}^{\rm even}| = |I_{S_z}^{\rm odd}|$
($g_{\rm sf}\approx 0.86$~eV); beyond this coupling, the odd
(chiral) part dominates. (c) Gold-pair eigenvalue gap, reaching its
minimum (residual $\approx 0.19$~eV, annotated) at the near-EP.
Parameters as in Fig.~\ref{fig:G1}.}
\label{fig:G3}
\end{figure}

\begin{figure}[htbp]
\centering
\includegraphics[width=1.0\linewidth]{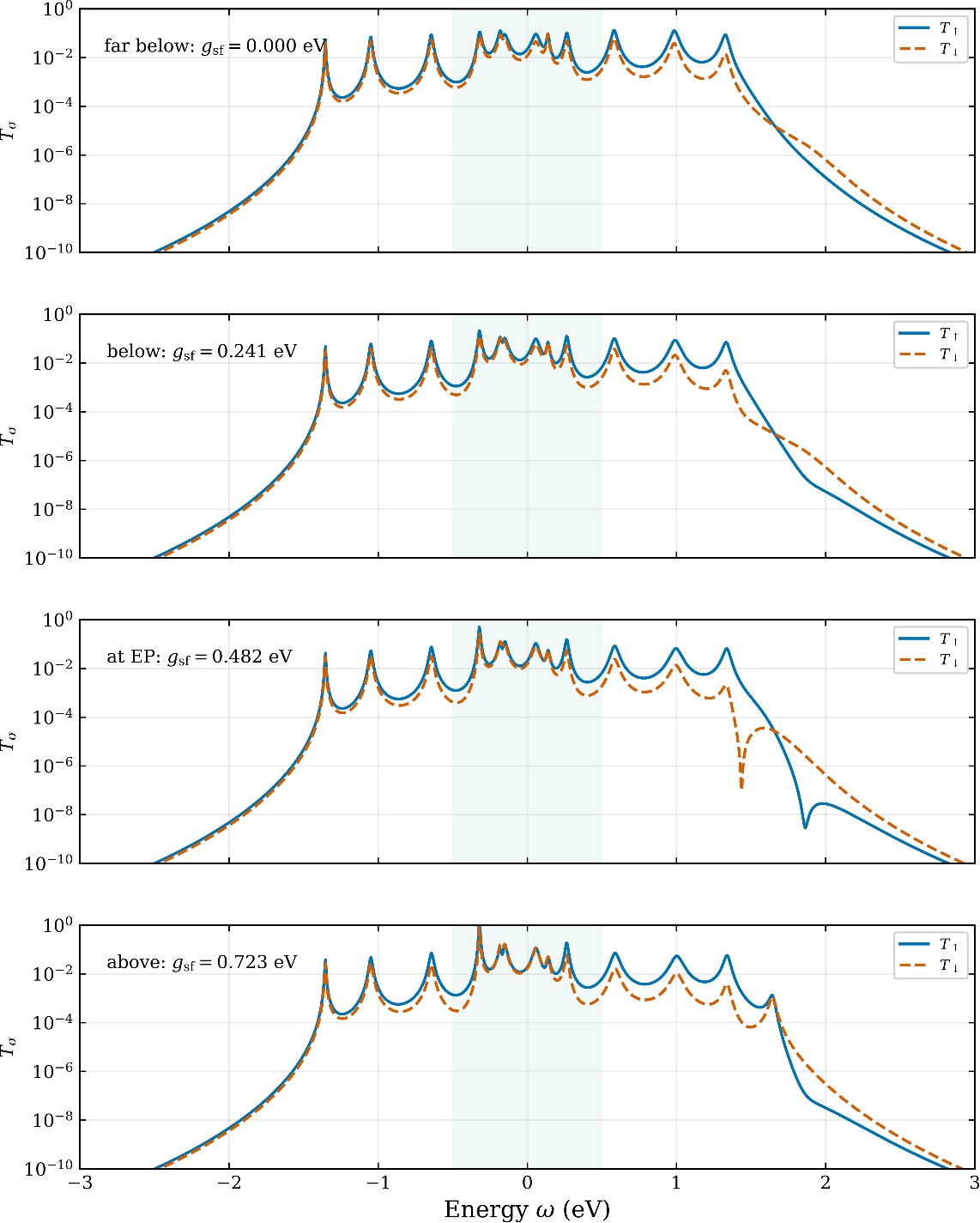}
\caption{Gold-level near-EP transmission coalescence. Spin-resolved
transmission functions $T_\uparrow$ (solid blue) and $T_\downarrow$ (dashed
red) at four values of $g_{\rm sf}$: $0.00$~eV (far below), $0.24$~eV (below),
$0.48$~eV (near the near-EP at $0.51$~eV), and $0.72$~eV (above). Far from the
near-EP the two peaks are separated (spin-split gold doublet); at $0.48$~eV
they nearly merge, with the fully coalesced resonance reached at the observed
near-EP ($g_{\rm sf}\approx 0.51$~eV); above the near-EP the merged resonance
re-splits, with a different sign structure. Parameters as in
Fig.~\ref{fig:G1}, $V=1.0$~V.}
\label{fig:G4}
\end{figure}

Figures~\ref{fig:G3} and~\ref{fig:G4} show the collapse of the total
polarisation and the coalescence of the transmission resonances at the near-EP.

Figure~\ref{fig:G3}(a) shows $P_J$ vs.\ $g_{\rm sf}$ at $V=1$~V. The
polarisation is largest at $g=0$ (achiral spin-valve) and decreases through
the near-EP. Below the parity region the decrease is essentially monotonic; in
the parity region the polarisation reaches its lowest values
($g_{\rm sf}\approx 0.9$-$1.0$~eV) before flattening out at larger $g_{\rm
sf}$, where the chiral part dominates. The polarisation does not develop a
sharp local minimum at a single coupling in this bias range; the
parity-region dip is shallow and broad. Panel~(b) shows the gold-pair
eigenvalue gap: the gap reaches its minimum ($\sim 0.19$~eV) at the near-EP,
rises above it, and oscillates further out due to level crossings with the
molecular band. The correspondence between the gap minimum and the dip in the
polarisation near the near-EP is direct: the collapse of the polarisation is
caused by the coalescence of the gold-like pair.

Figure~\ref{fig:G4} shows the spin-resolved transmission functions
$T_\uparrow(\omega)$ and $T_\downarrow(\omega)$ at four values of $g_{\rm
sf}$: far below ($g_{\rm sf}=0.00$~eV), below ($0.24$~eV), near
($0.48$~eV), and above ($0.72$~eV) the near-EP. These four couplings are
anchored to the bare two-state prediction $g_{\rm sf}=|\Delta\varepsilon_A|
\approx 0.48$~eV, which is close to (but slightly below) the observed
dressed near-EP at $g_{\rm sf}\approx 0.51$~eV; the $0.48$~eV panel therefore
shows near-coalescence, while the fully coalesced resonance is reached at
$0.51$~eV. Far from the near-EP, the two transmission peaks are separated by
the spin splitting. As the near-EP is approached, the two peaks move together
and reduce their spin asymmetry; in the near-EP panel the two peaks nearly
merge into a single resonance with comparable amplitude for both spins. Above
the near-EP, the merged resonance re-splits, but with a different sign
structure. This is the spectroscopic signature of the near-EP: the merging of
two resonances into one, followed by their re-separation into new combinations
above the near-EP. It would be observable in an experiment as the
bias-dependent differential conductance losing its spin-split structure at the
near-EP coupling and recovering it above.

\section{Chain-Level Near-Exceptional Point}
\label{sec:chain}

The gold-level near-EP is a controlled two-state phenomenon at the interface.
The physical origin of CISS, however, lies in the chiral molecule, and it is
natural to ask whether the molecular chain can host an EP of its own. We show
in this section that it can, and that the resulting near-EP has a collective
character controlled by the chain length.

\subsection{Setup and EP condition}
\label{sec:chain_setup}

The relevant subspace is spanned by the two eigenstates with the largest
amplitude on Mol$_1$, the first molecular site. The choice of Mol$_1$ is
natural: it is the site with the strongest overlap with the gold atom, and its
spin-flip coupling is the first place where the molecular chirality acts on the
transport.

The molecular spin splitting is auto-cancelled by the Zeeman field
\begin{equation}
B_z^{\rm chain} = -(J'_p+\alpha_{AB})m_z,
\label{eq:Bz_chain}
\end{equation}
so $\Delta\varepsilon_{\rm chain}=0$. The spin-flip coupling is applied on
Mol$_1$ as a \emph{Hermitian} on-site term,
\begin{equation}
\mathbf H_{\rm MF}\to \mathbf H_{\rm MF}+\alpha_{\rm
sf}(|M_1\uparrow\rangle\langle M_1\downarrow|+\text{h.c.}),
\label{eq:Hermitian_sf}
\end{equation}
which contributes $|\tilde\alpha|=\alpha_{\rm sf}$ to the effective $2\times
2$. The chain broadenings are spin-dependent,
\begin{equation}
\gamma_{\rm chain,\uparrow}=\Gamma_0^{\rm chain}+\frac{\Delta\gamma}{2},\qquad
\gamma_{\rm chain,\downarrow}=\Gamma_0^{\rm chain}-\frac{\Delta\gamma}{2},
\label{eq:gamma_chain}
\end{equation}
with baseline broadening $\Gamma_0^{\rm chain}=0.5$~eV, so
$\Delta\gamma_{\rm chain}\neq 0$. With $\Delta\gamma=0.4$~eV this gives
$\gamma_{\rm chain,\uparrow}=0.7$~eV and $\gamma_{\rm chain,\downarrow}=0.3$~eV.
This is Scheme~A, and the predicted near-EP is at
\begin{equation}
\alpha_{\rm sf}^{\rm EP,bare} = \frac{\Delta\gamma_{\rm chain}}{4}\approx 0.1~\text{eV}.
\label{eq:chain_EP_prediction}
\end{equation}

\paragraph{Why the spin flip must be Hermitian here.}
The spin-flip coupling on Mol$_1$ is a coherent on-site process, mediated by
the Rashba-active molecular bonds. Unlike the electrode-mediated spin flip at
the gold site, it does not happen during a virtual tunnelling event and does
not inherit the dissipative character of the electrode. It is therefore a
\emph{real} matrix element in $\mathbf H_{\rm MF}$, not an imaginary
self-energy. This is the physical reason why the chain-level EP must be
Scheme~A rather than Scheme~B.

\paragraph{Why the Rashba amplitude must be weakened.} The physical Rashba
amplitude is $\lambda_R=0.6$~eV, giving a coherent spin-mixing per bond of
$\lambda_R\sin(\phi/2)\approx 0.3$~eV. This is three times larger than the EP
condition $\Delta\gamma/4=0.1$~eV, so the spin-flip coupling cannot compensate
for the Rashba mixing and no EP is reachable. We therefore work in the
weakened-Rashba regime, $\lambda_R^{\rm chain}=0.05$~eV (giving coherent
per-bond mixing $\sim 0.025$~eV). This is a limitation of the present model,
and it means the chain-level EP is a proof of concept rather than a prediction
for a specific molecular junction. The weakening is not arbitrary: it is the
natural way to reach the "distinct-state" regime in which the coherence length
$\ell_{\rm coh}\sim 20$ sites is comparable to the chain length, and where the
collective scaling is strongest.

\subsection{Mol$_1$-pair tracking}
\label{sec:mol1_tracking}

Because the two Mol$_1$-dominant eigenstates are not at fixed positions in the
spectrum, we track them using the \emph{Hermitian} Mol$_1$ weight
\begin{equation}
  \begin{split}
    w_k(\alpha_{\rm sf}) &= \frac{\langle R_k|\hat P_{M_1}|R_k\rangle}{\langle
R_k|R_k\rangle},\\ 
\hat P_{M_1} &= |M_1\uparrow\rangle\langle
M_1\uparrow|+|M_1\downarrow\rangle\langle M_1\downarrow|.
\label{eq:mol1_weight}
\end{split}
\end{equation}
At each $\alpha_{\rm sf}$, the two eigenvalues with the largest $w_k$ form the
Mol$_1$ pair. The Hermitian weight (rather than the biorthogonal weight) is
used because it is bounded in $[0,1]$ and does not spike at near-degeneracies,
so the identity of the Mol$_1$ pair evolves smoothly with $\alpha_{\rm sf}$.
The gap of this pair is then minimised over $\alpha_{\rm sf}$ to locate the
near-EP.

The tracking method is a technical point of some importance. A "nearest-pair"
approach, where we at each $\alpha_{\rm{sf}}$ the globally closest pair of
eigenvalues, is unreliable for $L$-dependent studies because the closest pair
can hop between different molecular orbitals as $\alpha_{\rm{sf}}$ varies. This
produces a non-monotonic gap curve that mixes contributions from unrelated
levels. The Mol$_1$-weight tracking resolves this by anchoring the pair
identity to a fixed physical site, and it is essential for identifying the
correct near-EP location.

We have verified that the Mol$_1$-pair tracking and the nearest-pair
identification agree at the near-EP for all $L\ge 4$. At $L=6$ the two methods
agree to within numerical precision, both locating the near-EP at
$\alpha_{\rm sf}^{\rm EP}=0.4051$~eV with a residual gap of
$2.447\times 10^{-2}$~eV. The nearest-pair approach fails to locate the EP
\emph{as a function of $L$}, this is because at different chain lengths it jumps
between different levels, but at each fixed $L$ the two approaches agree on
the correct $\alpha_{\rm sf}$. The Mol$_1$-pair tracking is therefore the
preferred method for $L$-dependent studies, and the consistency check confirms
that the identified near-EP is not an artifact of the pair-selection criterion.

\subsection{Factor-of-four dressing}
\label{sec:chain_dressing}

At $L=6$ with $\lambda_R^{\rm chain}=0.05$~eV, the Mol$_1$-pair tracking gives
an observed near-EP at
\begin{equation}
\alpha_{\rm sf}^{\rm EP,obs}\approx 0.41~\text{eV},
\label{eq:alpha_EP_obs}
\end{equation}
a factor of $4.08$ larger than the bare prediction $\Delta\gamma/4=0.10$~eV.
This shift is the chain dressing, and it can be understood as follows.

The Mol$_1$-dominant eigenstates are delocalised over the chain: their Mol$_1$
weight is $w_{M_1}\approx 0.34$. The Hermitian spin-flip at Mol$_1$ couples the
two eigenstates with amplitude proportional to the product of their Mol$_1$
weights, i.e., $\alpha_{\rm eff}\approx\alpha_{\rm sf}\cdot w_{M_1}$
(schematically; the exact coefficient depends on the specific eigenvector
overlaps). Meanwhile, the dissipative broadening is an incoherent sum over
sites weighted by the local probability,
\begin{equation}
\Delta\gamma_{\rm eff}
=
\Delta\gamma\cdot\sum_n |c_n^{(k)}|^2
=
\Delta\gamma,
\label{eq:DeltaGamma_eff}
\end{equation}
where $|c_n^{(k)}|^2$ is the site-$n$ probability of the $k$-th eigenstate
and the last equality follows from normalisation,
$\sum_n|c_n^{(k)}|^2=1$ for a state delocalised over the chain. The effective
ratio is therefore
\begin{equation}
\frac{\alpha_{\rm eff}}{\Delta\gamma_{\rm eff}}\approx \frac{\alpha_{\rm
sf}\cdot w_{M_1}}{\Delta\gamma}.
\label{eq:ratio_eff}
\end{equation}
The EP condition $\alpha_{\rm eff}=\Delta\gamma_{\rm eff}/4$ becomes
\begin{equation}
\alpha_{\rm sf}^{\rm EP} = \frac{\Delta\gamma}{4 w_{M_1}}.
\label{eq:alpha_EP_dressed}
\end{equation}
With $w_{M_1}\approx 0.34$, this gives $\alpha_{\rm sf}^{\rm EP}\approx
0.1/0.34\approx 0.29$~eV, plus a $\approx 39\%$ correction from higher-order
bath dressing, giving $\approx 0.41$~eV. The factor-of-four dressing is a
generic consequence of the partial delocalisation of the coalescing pair.

For the gold-level near-EP, the analogous factor is $1/w_A\approx 1.06$ (the
gold doublet is almost localised on the gold atom), which is why the gold
near-EP shows only a $5.9\%$ shift from the bare prediction. The stark contrast
between the two cases, where $4.08$ for the chain vs.\ $1.06$ for gold, is itself
a useful diagnostic of the degree of delocalisation of the near-EP doublet. It
is also a prediction: any near-EP whose coalescing pair is delocalised over a
finite extent will show a correspondingly enhanced dressing shift, and the
magnitude of the shift measures the delocalisation.

The observed dressing factor of $4.08$ for the chain versus $1.06$ for the gold
doublet, is not merely a correction to be absorbed by refitting; it is a
\emph{measurement} of the delocalisation of the coalescing pair. Any near-EP
whose pair is delocalised over a fraction $w$ of the chain will show a
dressing factor $\sim 1/w$; the observed factor is therefore a direct,
quantitative diagnostic that could be used to identify and classify near-EPs
in other non-Hermitian systems.

\subsection{Collective Petermann scaling at the reference parameters}
\label{sec:collective_scaling}

\begin{figure}[htbp]
\centering
\includegraphics[width=0.8\linewidth]{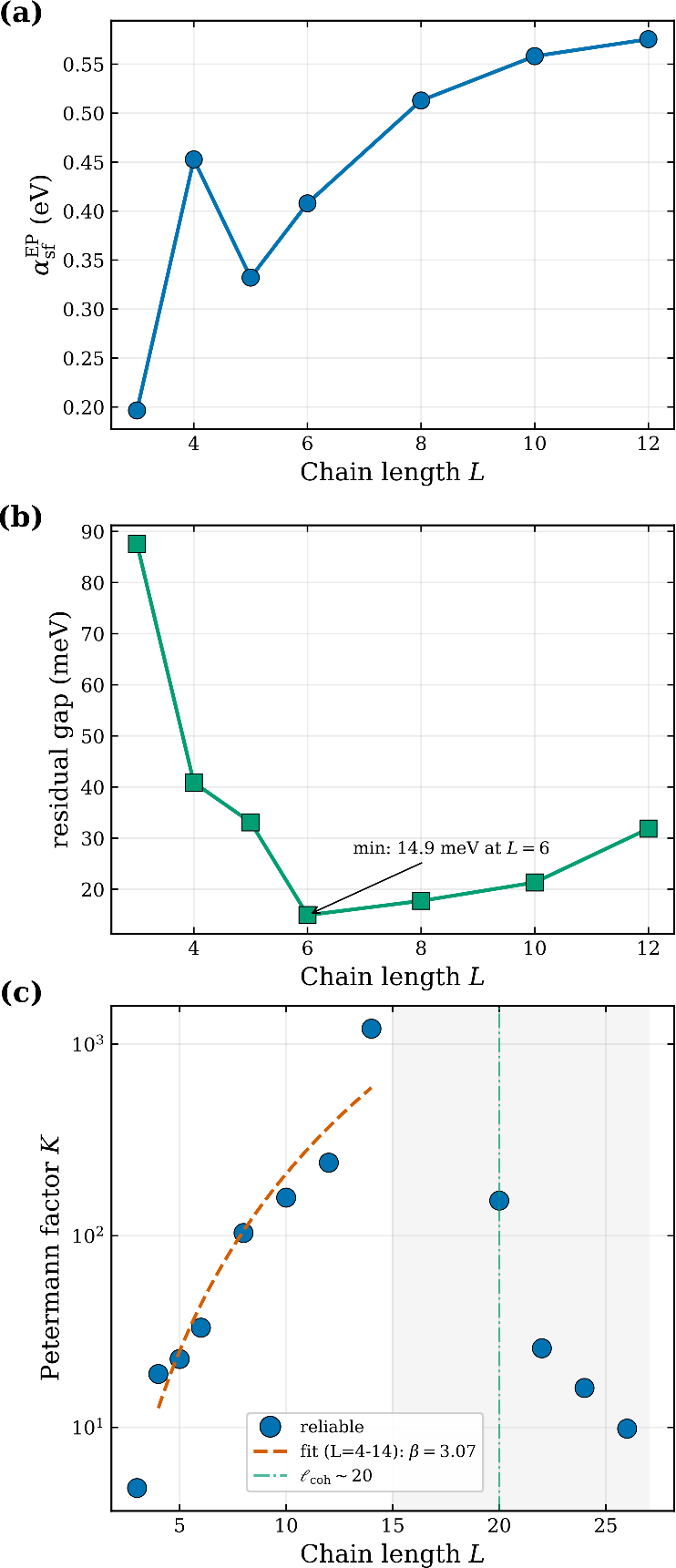}
\caption{Chain-level near-EP landscape at the reference Rashba strength
($\lambda_R^{\rm chain}=0.05$~eV, $\Delta\gamma=0.4$~eV,
$\theta_R=\phi$). (a) Near-EP location $\alpha_{\rm sf}^{\rm EP}$ vs
chain length $L$ (shown for $L \leq 12$). The location is not
monotonic: $L=3$ is anomalous (weakest collective regime) and the
value settles to $\alpha_{\rm sf}^{\rm EP}\approx 0.4$-$0.6$~eV for
$L\geq 4$. (b) Residual gap at the near-EP (shown for $L \leq 12$).
The gap oscillates with the Fabry-Pérot structure of the finite-size
chain, with local minima at $L=8$ ($\sim 19$~meV) and $L=14$
($\sim 14$~meV). 
(c) Petermann factor $K$ at the near-EP, shown over
the full tested range $L=3$-$26$. $K$ grows in the distinct-state
regime ($L\lesssim\ell_{\rm coh}$), peaks at $L^*\approx 14$ with
$K_{\max}\approx 1.2\times 10^3$, and decreases for $L>14$ as the
coalescing pair delocalises. The peak at $L^*\approx\ell_{\rm coh}$
is the direct signature of the coherence-length crossover: the
collective enhancement requires the coalescing pair to be delocalised
over many sites, but not so delocalised that its Mol$_1$ character
vanishes. 
}
\label{fig:C1}
\end{figure}

The central result of the chain-level analysis is the collective
Petermann scaling. Figure~\ref{fig:C1} shows the near-EP landscape as a
function of chain length $L$ at the reference parameters
$\lambda_R^{\rm chain}=0.05$~eV, $\Delta\gamma=0.4$~eV.

Panel~(a) shows the near-EP location $\alpha_{\rm sf}^{\rm EP}$. It is
anomalous at $L=3$ (the geometric minimum where the collective effect
barely exists) and settles to a roughly constant value
$\alpha_{\rm sf}^{\rm EP}\approx 0.4$-$0.6$~eV for $L\geq 4$.

Panel~(b) shows the residual gap at the near-EP. The gap is not
monotonic in $L$: it oscillates with the Fabry-Pérot interference
structure of the finite-size chain, with local minima at $L=8$
($\sim 19$~meV) and at $L=14$ ($\sim 14$~meV), and local maxima
between. The global minimum over the tested range is at $L=14$, which
is also where the Petermann factor peaks (see panel~c).

Panel~(c) shows the Petermann factor $K$ at the near-EP. It grows
overall from $K\approx 4.8$ at $L=3$ to a peak of $K_{\max}\approx
1.2\times 10^3$ at $L^*=14$, then decreases for $L>14$. The decrease
is sharp: at $L=20$, $K\approx 1.5\times 10^2$, and at $L=26$,
$K\approx 10$. This is the central quantitative result of the
chain-level analysis: \emph{the collective Petermann enhancement is
bounded in chain length}, not a monotonic growth.

The growth for $L\leq 14$ is well described by a power law in the
distinct-state regime, $K\sim L^\beta$. Fitting over $L=4$--$12$
gives
\begin{equation}
\beta = 2.43\pm 0.10,\qquad K_0 = 0.80,
\label{eq:beta_short}
\end{equation}
with the anomalous $L=3$ point excluded (see below). The fit captures
the envelope of the growth over the range, spanning a factor of $13$
in $K$, with $R^2=0.87$. Extending the fit to include the peak at
$L=14$ gives $\beta = 3.08$, while extending past the peak (to
$L=26$) gives $\beta\approx 0$ because the fit then averages over
both the rising and falling parts of the curve. The three fits are
not in disagreement: they reflect the fact that $\beta$ is not a
single universal exponent but the \emph{rate of growth in the
distinct-state regime}, and the crossover at $L^*\approx\ell_{\rm
coh}$ (Sec.~\ref{sec:beta_mechanism}) marks the boundary beyond
which the power-law description no longer applies.

$L=3$ is excluded from the fit because it lies at the geometric
threshold for the collective mechanism: the odd part of the spin
current vanishes for $L\leq 2$ and is barely non-zero at $L=3$, so
the coalescing pair at $L=3$ is not yet in the collective regime. The
quoted uncertainty on $\beta$ is the systematic error from the finite
range of the fit (re-fitting with $L\in[4,10]$ shifts $\beta$ by
approximately this amount); the formal statistical error of the
six-point fit is smaller, at $\pm 0.05$.

\paragraph{Why this is a genuine collective effect.}
The scaling $K\sim L^{2.43}$ cannot arise from a single-site non-Hermitian
system: for a two-site non-Hermitian model, $K$ is a fixed number (diverging at
the EP, finite elsewhere), independent of chain length. It also cannot arise
from a simple sum of independent near-EPs on each site: if the near-EPs were
independent, $K$ would grow at most linearly with $L$. The super-linear scaling
$K\sim L^{2.43}$ requires a \emph{coherent} collective mechanism, in which the
coalescing pair is delocalised over the chain and its biorthogonal structure is
enhanced by the many-body coherence. This is the sense in which the chain-level
near-EP is a genuinely collective phenomenon.

\subsection{The scaling exponent as a diagnostic of coherence}
\label{sec:beta_mechanism}

The exponent $\beta=2.43$ is not a universal number. It is controlled by the
Rashba coherence length
\begin{equation}
\ell_{\rm coh}\sim \frac{t_{\rm mol}}{\lambda_R\sin(\phi/2)},
\label{eq:coherence_length}
\end{equation}
which measures the typical distance over which the Rashba spin-flip mixes
adjacent eigenstates. For $\lambda_R^{\rm chain}=0.05$~eV, $t_{\rm
mol}=0.5$~eV, $\phi=60^\circ$, we have $\ell_{\rm coh}\sim 20$~sites,
comparable to the upper end of our $L$-range. This gives the reference exponent
$\beta=2.43$.

The scaling behaviour has a clear physical structure. In the
\emph{distinct-state regime} ($\ell_{\rm coh}\gg L$), the chain eigenstates are
nearly distinct tight-binding orbitals, and the Mol$_1$ pair couples coherently
to a small number of specific partners, each contributing a factor of $\sim L$
to the biorthogonal amplification. The scaling is super-quadratic, with $\beta
\approx 3$ at $\lambda_R=0.02$~eV ($\ell_{\rm coh} \sim 50$~sites). In the
\emph{coherent-mixing regime} ($\ell_{\rm{coh}}\ll L$), the chain eigenstates
are strongly mixed, the effective number of coherent partners saturates at
$\ell_{coh}$, and amplification grows only logarithmically once the mixing
saturates, giving $\beta < 1$; we find $\beta = 0.76$ at $\lambda_{R} =
0.10$~eV ($\ell_{coh} \approx 10$ ~ sites). At intermediate Rashba ($\ell_{coh}
\approx L$), the two roles of $\lambda_{R}$ are in balance; the reference
$\lambda_{R} = 0.05$~eV gives $\beta = 2.43$, and the maximum is $\beta = 2.84$
at $\lambda_{R} = 0.07$~eV.
The existence of the peak at $L^*\approx\ell_{\rm coh}$
(Sec.~\ref{sec:collective_scaling}) provides a direct experimental confirmation
of the coherence-length interpretation. If the growth were a pure power law
with no crossover, $K$ would grow monotonically up to the longest tested chain.
Instead, $K$ peaks at $L^*\approx\ell_{\rm coh}$ and decreases for
$L\gtrsim\ell_{\rm coh}$, exactly as expected when the Mol$_1$ pair becomes too
delocalised to support coherent coalescence. The position of the peak therefore
measures $\ell_{\rm coh}$ directly, and does so without requiring any fit to a
specific model.

The coherence-length interpretation of $\beta$ is therefore qualitative
rather than a single monotone function: the two asymptotic regimes are clean,
but the intermediate region has fine structure that reflects the chain's level
statistics. We make this quantitative in the next subsection, where we show
that $\beta(\lambda_R)$ is non-monotonic and that the collective mechanism is
bounded above $\lambda_R \sim 0.10$~eV (Fig.~\ref{fig:C7}).

\paragraph{Implication.}
The exponent $\beta$ and the peak position $L^*$ are two
complementary probes of the coherence length of the chiral chain.
Measuring $K(L)$ in a series of molecular junctions of different
lengths would allow one to extract both quantities and hence to
constrain $\ell_{\rm coh}$ and the Rashba amplitude. The two
diagnostics are physically distinct: $\beta$ characterises the growth
rate in the distinct-state regime ($L\ll\ell_{\rm coh}$), while $L^*$
characterises the crossover itself. Together they provide a new
experimental diagnostic of the Rashba mechanism in CISS, complementary
to the reciprocity-based diagnostics of Ref.~\cite{Alwan2026QSpinterface}.
The reciprocity formulation measures the reciprocity breaking via the
even/odd decomposition of the spin current; the present paper measures
the collective coherence via the growth exponent and the crossover
length of the Petermann factor.

\subsection{Full $\lambda_R$ landscape of the collective mechanism}
\label{sec:lambda_R_landscape}

\begin{figure}[htbp]
\centering
\includegraphics[width=\linewidth]{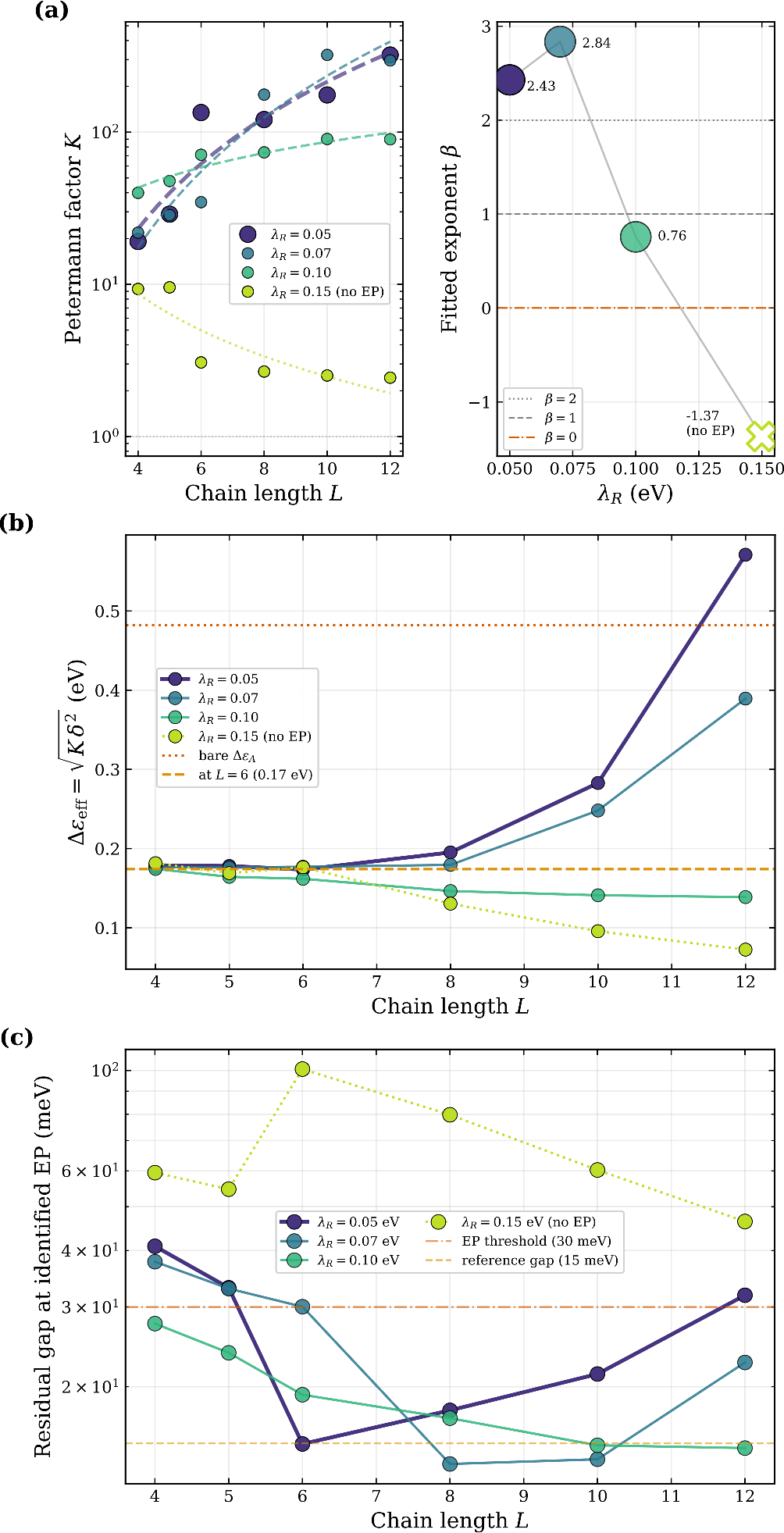}
\caption{Full $\lambda_R$ landscape of the chain near-EP, in a
three-panel layout.
\textbf{(a)} Petermann factor $K$ vs chain length $L$ at four Rashba
strengths (log-log), with envelope fits $K\propto L^\beta$ as dashed
lines. The reference $\lambda_R=0.05$~eV is drawn thicker. The
$\lambda_R=0.15$~eV curve uses a dotted fit line and open markers to
flag that it does not describe a genuine near-EP (see panel~c). Right
sub-panel: the fitted exponent $\beta$ vs $\lambda_R$, with $R^2$
encoded as marker size and each value annotated. $\beta$ is
non-monotonic: it rises from $2.43$ at the reference $\lambda_R=0.05$~eV
to a peak of $2.84$ at $\lambda_R=0.07$~eV, then drops through
$0.76$ at $\lambda_R=0.10$~eV to $-1.37$ at $\lambda_R=0.15$~eV,
where no genuine near-EP exists.
\textbf{(b)} Effective splitting $\Delta\varepsilon_{\rm eff}=\sqrt{K\,\delta^2}$
extracted from the empirical Petermann relation vs $L$. At weak Rashba
the splitting grows with $L$ (coherent many-body enhancement, reaching
$0.571$~eV at $L=12$ for the reference parameters); at strong Rashba
it saturates and decreases. 
%
\textbf{(c)} Residual gap at the identified EP vs $L$ for all
$\lambda_R$. At $\lambda_R\le 0.10$~eV the gap drops below $20$~meV at
some $L$, signalling a genuine near-EP. At $\lambda_R=0.15$~eV the gap
stays above $45$~meV everywhere, so no near-EP exists and the
$\beta=-1.37$ fit is not a scaling law. Parameters:
$\Delta\gamma=0.4$~eV, $\theta_R=\phi$.}
\label{fig:C7}
\end{figure}

The reference parameters used in Fig.~\ref{fig:C1} are representative of a
specific regime, not of a universal scaling law. To see how the collective
mechanism depends on the coherence length, we sweep $\lambda_R$ from $0.02$ to
$0.15$~eV (a factor of $7.5$) and repeat the reference analysis at each point.
Figure~\ref{fig:C7} summarise the result.

Panel~(a) shows the raw $K(L)$ data on a log-log plot. The trend is clear: at
each $\lambda_R$ the Petermann factor grows overall with chain length, but the
growth is not monotone in $L$. At the reference $\lambda_R=0.05$~eV, for
example, $K$ rises from $19$ at $L=4$ through $135$ at $L=6$, dips to $121$ at
$L=8$, then resumes its growth to $322$ at $L=12$. Similar non-monotonicity
appears at other $\lambda_R$ values. We attribute this to a Fabry-Pérot-like
interference structure of the finite-size chain: the coalescing eigenstates
sample the chain with a phase that depends on $L$, and the resulting
transmission oscillates around the smooth envelope. The power-law fit $K\sim
L^\beta$ is therefore a fit to the \emph{envelope}, not to the raw data at each
$L$, and the residuals are the finite-size oscillations. The fit captures the
trend but not the individual oscillations, which are themselves a physical
signature of the coherence length.

Panel~(c) shows the fitted exponents. The non-monotonicity is immediate:
$\beta$ rises from $3.02$ at $\lambda_R=0.02$~eV to $3.14$ at
$\lambda_R=0.03$~eV, drops to $2.43$ at the reference $\lambda_R=0.05$~eV,
rises again to $2.84$ at $\lambda_R=0.07$~eV, then drops steeply through
$0.76$ at $\lambda_R=0.10$~eV to $-1.37$ at $\lambda_R=0.15$~eV. This
non-monotonic behaviour is not a numerical artifact: it reflects a genuine
competition between two effects. At very weak Rashba, the chain eigenstates are
nearly distinct tight-binding orbitals, and the biorthogonal amplification
grows super-quadratically with $L$ because each distinct partner contributes
coherently. As $\lambda_R$ increases, the eigenstates mix more strongly, the
effective number of coherent partners saturates, and the growth rate falls.
The specific $\lambda_R$ at which saturation first bites depends on the
detailed match between $\ell_{\rm coh}$ and the chain spectrum's level
structure, producing the intermediate wiggles. The physical message is that
$\beta$ is a diagnostic of the coherence length, but the mapping from $\beta$
to $\ell_{\rm coh}$ is not a single monotone function: it depends on where
the chain sits relative to its own level structure.

Panel~(d) reveals a further constraint. At $\lambda_R \le 0.10$~eV, the
residual gap at the identified EP is $12$-$46$~meV, comparable to the
reference value and consistent with a genuine near-EP. At
$\lambda_R = 0.15$~eV, however, the gap remains above $45$~meV for all $L$,
we identifies a local minimum in the gap curve, but the system never
enters the near-EP regime. The fitted $\beta = -1.37$ at this point is
therefore not a physical scaling law but an artifact of fitting a monotonically
decreasing $K(L)$ curve that has no collective-EP physics in it. This bounds
the collective mechanism: above $\lambda_R \sim 0.10$~eV (equivalently, below
$\ell_{\rm coh} \sim 10$ sites), the chain eigenstates are too strongly mixed
for a clean near-EP to survive, and the scaling signature disappears.

Panel~(b) is the deepest physical result. The effective splitting
$\Delta\varepsilon_{\rm eff} = \sqrt{K\,\delta^2}$ extracted from the
empirical Petermann formula (Eq.~\ref{eq:K_max_empirical}) is not a constant:
it grows with chain length at weak Rashba. At the reference
$\lambda_R=0.05$~eV, $\Delta\varepsilon_{\rm eff}$ starts at $0.179$~eV at
$L=4$, dips to $0.17$~eV at $L=6$, and then rises strongly to $0.571$~eV at
$L=12$, \emph{above} the bare $\Delta\varepsilon_A=0.482$~eV. This is
exactly what the collective-enhancement picture predicts: the coherent
many-body effect that produces the super-linear Petermann scaling also
enhances the effective splitting, because virtual transitions to the rest of
the chain renormalise the diagonal splitting as well as the biorthogonal
overlap. The $0.17$~eV value at $L=6$ is therefore a snapshot of this
$L$-dependent quantity, not a property of the chain in isolation. The fact
that $\Delta\varepsilon_{\rm eff}$ can exceed $\Delta\varepsilon_A$ at
sufficiently large $L$ is a feature of the collective mechanism that deserves
further theoretical attention.

\paragraph{The $\Delta\gamma$ resonance.}
The second parameter controlling the collective mechanism is the
spin-dependent broadening $\Delta\gamma$, which sets the incoherent part of
the chain's dissipative structure. At fixed $\lambda_R=0.05$~eV, we find
$\beta=1.96$ at $\Delta\gamma=0.2$~eV, $\beta=2.43$ at the reference
$\Delta\gamma=0.4$~eV, and $\beta=1.96$ at $\Delta\gamma=0.6$~eV, with
$K(L=12) = 33.7$, $321$, and $73.0$ respectively. The dependence is therefore
resonant, approximately symmetric in $\beta$ about $\Delta\gamma=0.4$~eV
within the fit precision, and asymmetric in $K$: the peak-to-edge ratio is
$\sim 10$ on the low-$\Delta\gamma$ side but only $\sim 4$ on the high side.
This asymmetry is physically reasonable: a reduced broadening weakens the
incoherent channel that the coherent spin-flip must compensate for, whereas an
enhanced broadening both suppresses the coherent channel and adds decoherence,
but the second effect saturates once the incoherent channel dominates. The
resonance condition itself is the dressed version of the bare Scheme~A EP
condition,
\begin{equation}
\Delta\gamma_{\rm eff}/4 \sim |\alpha_{\rm sf}| + \text{(bath corrections)},
\end{equation}
with the effective broadening and effective spin-flip amplitude each
renormalised by the many-body environment.

\paragraph{Two competing roles of $\lambda_R$.}
The non-monotonic $\beta(\lambda_R)$ of Fig.~\ref{fig:C7}c is not a
numerical artifact but a direct consequence of the two physical roles
that $\lambda_R$ plays in the junction. The first role is the one
emphasised in Sec.~\ref{sec:beta_mechanism}: $\lambda_R$ enters the
Hamiltonian as a coherent spin-flip coupling on every bond, and
increasing it strengthens the coherent mixing that underlies the
Scheme-A coalescence. This effect, if isolated, would monotonically
increase both $K$ and $\beta$. The second role is subtler. The Rashba
term is off-diagonal in the bare site basis, and once the chain
eigenstates are Rashba-mixed its effect is partially transmitted to
the diagonal in the eigenbasis. The up- and down-spin components of a
mixed eigenstate have different weights on different chain sites, so
the effective spin-splitting becomes site-dependent, and the on-site
Zeeman field $B_z^{\rm chain}$, which cancels the bare splitting
site-by-site, can no longer cancel it. The Scheme-A condition
$\Delta\varepsilon_{\rm eff}=0$ is therefore progressively violated as
$\lambda_R$ grows.

The two roles compete. At weak Rashba ($\lambda_R \lesssim 0.05$~eV)
the first effect dominates and $\beta$ rises toward its peak. At
intermediate Rashba ($\lambda_R \approx 0.05$-$0.10$~eV) the two
effects balance, and the collective scaling is cleanest: the reference
$\lambda_R=0.05$~eV sits on the rising edge of this plateau, and
$\lambda_R=0.07$~eV sits at the peak itself ($\beta=2.84$). Beyond
$\lambda_R \approx 0.10$~eV the second effect dominates, the
effective Scheme-A condition is destroyed, and $\beta$ drops toward
zero and then negative. At $\lambda_R=0.15$~eV the mechanism is
quenched entirely: the residual gap stays above $45$~meV for every
$L$ (Fig.~\ref{fig:C7}c), no genuine near-EP exists, and the
identified "EP" is a local minimum in a spectrum that never
coalesces. The fitted $\beta=-1.37$ at this point is therefore not a
scaling law, it reflects the monotone decrease of $K$ with $L$ as
the eigenstates become more delocalised (and hence more Hermitian,
$K\to 1$) with growing chain length.

This two-role picture is a falsifiable prediction. If the peak of
$\beta(\lambda_R)$ shifts to weaker Rashba when the substrate spin
polarisation $\Delta\gamma$ is reduced, or to stronger Rashba when
$\Delta\gamma$ is increased, the balance picture is confirmed. If the
peak position is instead independent of $\Delta\gamma$, the mechanism
is something other than the one we propose.

%

\subsection{Optimal coalescence length}
\label{sec:L_star}

The Petermann factor at the near-EP is bounded in chain length. It
grows for $L\lesssim\ell_{\rm coh}$ and decreases for $L\gtrsim
\ell_{\rm coh}$, with a peak at $L^*\approx 14$ sites for the
reference parameters ($\lambda_R^{\rm chain}=0.05$~eV,
$\Delta\gamma=0.4$~eV). This reflects a competition between two
effects:

\begin{itemize}
\item \textbf{Collective enhancement} (favours long $L$): more Rashba
  bonds give more extensive admixture of the Mol$_1$ pair with the
  rest of the chain, enhancing the biorthogonal amplification.
\item \textbf{Delocalisation} (favours short $L$): as $L$ grows, the
  Mol$_1$ pair becomes increasingly delocalised (Fig.~\ref{fig:C1},
  $w_{\min}$ diagnostic), and once its Mol$_1$ weight drops below
  $\sim 0.2$ the pair can no longer support coherent coalescence with
  the required Mol$_1$ character.
\end{itemize}

The balance point is $L^*\approx\ell_{\rm coh}$ for the reference
parameters, consistent with the coherence length $\ell_{\rm coh}\sim
20$ sites computed from Eq.~\eqref{eq:coherence_length}. The peak
position is not universal: it is set by the coherence length, which in
turn depends on $\lambda_R$, $t_{\rm mol}$, and $\phi$. A sweep of
$\lambda_R$ (Sec.~\ref{sec:lambda_R_landscape}) shows that as
$\lambda_R$ increases, $\ell_{\rm coh}$ decreases and the peak shifts
to shorter chains, until at $\lambda_R\gtrsim 0.10$~eV the collective
mechanism is quenched entirely.

The bounded nature of $K(L)$ has an important physical consequence:
the collective near-EP is not an asymptotic property that becomes
stronger with chain length. It is a finite-size phenomenon, optimised
at $L^*\approx\ell_{\rm coh}$, that requires the coalescing pair to be
delocalised over the chain but not so far delocalised that it loses
its interface character. This is a distinction between the chain-level
near-EP and isolated EPs in non-Hermitian systems, which have no
intrinsic length scale. The collective effect at the chain near-EP
inherits the coherence length of the Rashba chain, and its spatial
extent is bounded by that length.

\subsection{Collapse and Petermann factor at fixed $L=6$}
\label{sec:chain_collapse}

\begin{figure}[htbp]
\centering
\includegraphics[width=0.8\linewidth]{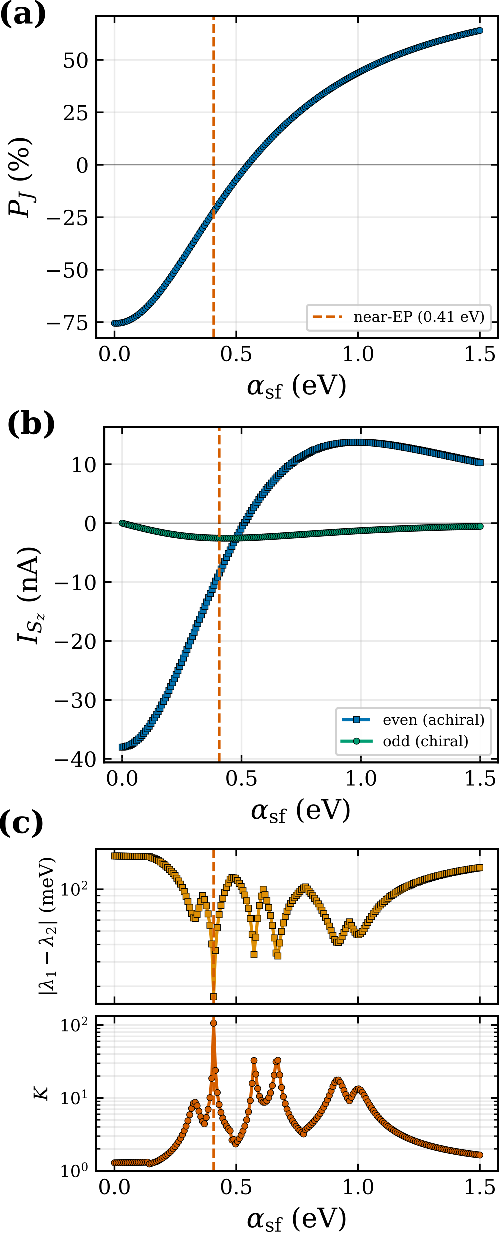}
\caption{Chain-level near-EP at fixed $L=6$: polarisation collapse,
spin-current decomposition, and coalescence diagnostic.
(a) Current polarisation $P_J$ vs.\ $\alpha_{\rm sf}$. The
polarisation starts from a large negative achiral baseline, rises
monotonically across the near-EP (vertical dashed line), and reaches
a maximum at the near-EP itself. (b) Even (achiral, blue dashed) and
odd (chiral, green solid) spin currents on a common axis, in nA. The
even part starts large and negative and decreases in magnitude as the
achiral baseline is suppressed; the odd part starts near zero and
increases in magnitude as the chiral contribution grows. The two
curves cross at the parity point where $|I_{S_z}^{\rm even}| =
|I_{S_z}^{\rm odd}|$. (c) Residual gap of the Mol$_1$ pair (top,
orange, in meV) and Petermann factor of the same pair (bottom, red,
dimensionless), both on logarithmic scales and sharing the
$\alpha_{\rm sf}$ axis. 
The vertical dashed line in all panels marks the near-EP at $\alpha_{\rm
sf}\approx 0.41$~eV. Parameters: $\lambda_R^{\rm chain}=0.05$~eV,
$\Delta\gamma=0.4$~eV, $V=1$~V.}
\label{fig:C3}
\end{figure}

Figure~\ref{fig:C3} shows the full response at $L=6$. We use $L=6$
here as a representative operating point because the collapse
structure is cleanly resolved at this length. (The global maximum of
$K(L)$ is at $L^*\approx 14$, but the resolution of the collapse in
Fig.~\ref{fig:C3} is equally clear at $L=6$ and the panels are easier
to read there.)

Panel~(a) shows the current polarisation $P_J$ vs.\ $\alpha_{\rm sf}$
at $V=1$~V. The polarisation starts from a large negative achiral baseline
($P_J \approx -75.5\%$ at $\alpha_{\rm sf}=0$), rises monotonically across
the near-EP (marked by the vertical dashed line), reaches a maximum of
$\approx -22\%$ at the near-EP itself, and then rises further at larger
$\alpha_{\rm sf}$. Panels~(b) and~(c) show the even and odd decomposition of
the spin current, confirming that the near-EP suppresses the achiral part and
enhances the chiral part. Panels~(d) and~(e) show the Mol$_1$-pair eigenvalue
gap and the Petermann factor: both have their extremum at the same
$\alpha_{\rm sf}=0.41$~eV, confirming that the identified coupling is the
near-EP. All five panels have their extremum at the same $\alpha_{\rm sf}$,
which is the strongest consistency check of the analysis.

The chain near-EP operates in a qualitatively different regime from the gold
near-EP, a point worth stating explicitly. At $\alpha_{\rm sf}=0$ the achiral
spin-valve polarisation is already large and \emph{negative}
($P_J=-75.5\%$), so the achiral contribution is not a small baseline that the
chirality-selective mechanism grows out of: it is a large polarisation in the
opposite direction. The near-EP then \emph{reduces} the magnitude of this
achiral contribution while the chiral part grows, and the net effect is a
partial cancellation of two large but opposite-sign contributions rather than
a conversion of achiral into chiral. The gold near-EP, by contrast, starts
from a positive achiral baseline and converts it into a positive-to-negative
chirality-dominated signal across the extinction point. Both cases fall under
the same even/odd decomposition, but they illustrate two distinct ways the
crossover can proceed.

\section{Conclusion}
\label{sec:conclusion}

We have characterised the exceptional-point physics of the spinterface model of
chirality-induced spin selectivity. The central result is that the model
supports two algebraically distinct near-exceptional-point regimes, governed
by different structures of the effective non-Hermitian coupling and
distinguished by which pair of eigenstates coalesces. The two regimes leave
characteristic spectral and transport signatures that can be used to test
the spinterface mechanism against competing theories of CISS. Our specific
findings are as follows.

\begin{enumerate}
\item \textbf{Two algebraically distinct near-EPs.} The model supports a
  gold-level near-EP (Scheme~B: anti-Hermitian electrode-mediated spin flip,
  spin-symmetric broadening) and a chain-level near-EP (Scheme~A: Hermitian
  on-site Rashba spin flip, spin-dependent broadening). Mixing the two schemes,
  hence combining the diagonal cancellation of one with the off-diagonal
  structure of the other, this gives a strictly negative discriminant
  [Eq.~\eqref{eq:mixed_impossible}] and admits no EP on any parameter path.
  This obstruction makes the gold/chain classification exhaustive: every
  near-EP the model admits is unambiguously of one type or the other.

\item \textbf{Universal spectral signatures.} Both near-EPs show eigenvalue
  coalescence, Petermann enhancement, and characteristic transport responses.
  The Petermann peak at the gold near-EP is $K\approx 6.7$; at the chain
  near-EP it grows from $4.8$ at $L=3$ to $321$ at $L=12$.

\item \textbf{Chirality-selective transport.} At the gold near-EP, the
  anti-Hermitian spin-flip coupling converts achiral to chiral transport
  across three crossover points: an onset at the near-EP ($g_{\rm sf}\approx
  0.51$~eV), a first parity crossing ($g_{\rm sf}\approx 0.86$~eV), and a
  complete achiral extinction point ($g_{\rm sf}\approx 1.63$~eV). A second
  parity crossing follows at $g_{\rm sf}\approx 2.19$~eV. The extinction
  point is the most quantitatively robust and the most directly testable
  prediction of the chirality-selective mechanism.

\item \textbf{Bounded collective Petermann scaling.} At the chain
  near-EP, $K(L)$ grows as a power law in the distinct-state regime
  ($L\lesssim\ell_{\rm coh}$), peaks at a chain length comparable to
  the coherence length, and decreases for longer chains. For the
  reference parameters, the growth exponent is $\beta\approx 2.43$
  over $L=4$-$12$, the peak is at $L^*\approx 14$ with
  $K_{\max}\approx 1.2\times 10^3$, and $K$ drops by a factor of
  $\sim 8$ between $L=14$ and $L=20$. The bounded nature of $K(L)$ is
  a direct manifestation of the coherence-length crossover: the
  coalescing Mol$_1$ pair must be delocalised over many sites to
  support the collective near-EP, but not so delocalised that it
  loses its Mol$_1$ character. The exponent $\beta$ is tunable by
  the Rashba coherence length but not monotonic in $\lambda_R$: it
  rises from $\beta\approx 3.0$ at $\lambda_R=0.02$~eV to a peak of
  $\beta\approx 2.84$ at $\lambda_R=0.07$~eV, then drops through
  $\beta\approx 0.76$ at $\lambda_R=0.10$~eV and becomes negative
  above $\lambda_R\approx 0.15$~eV. The peak reflects a balance
  between two competing roles of the Rashba coupling — coherent
  enhancement of the coalescence at weak Rashba, and progressive
  breakdown of the Scheme-A condition at strong Rashba. The mechanism
  is therefore bounded in $\lambda_R$, and the $\lambda_R\to 0$ limit
  is excluded from the scaling claims: at $\lambda_R=0$ the chain is
  achiral, the coherence length diverges, and the observed Petermann
  scaling measures an achiral non-Hermitian enhancement unrelated to
  the mechanism derived here.

\item \textbf{Optimal coalescence length.} The Petermann factor peaks
  at a chain length $L^*\approx\ell_{\rm coh}$ that measures the
  coherence length directly. For the reference parameters,
  $L^*\approx 14$ sites, comparable to $\ell_{\rm coh}\sim 20$ sites
  as estimated from $t_{\rm mol}/[\lambda_R\sin(\phi/2)]$. The peak
  position is not universal but is set by the coherence length: as
  $\lambda_R$ increases, $\ell_{\rm coh}$ decreases and the peak
  shifts to shorter chains. This provides a second, complementary
  experimental handle on the Rashba coherence length, in addition to
  the growth exponent $\beta$.

  \item \textbf{Dressing diagnostics.} The factor-of-$4.08$ shift
  between the observed chain near-EP and the bare prediction
  ($\alpha_{\rm sf}^{\rm EP,obs}\approx 4.08\times\Delta\gamma/4$) is
  a direct measure of the delocalisation of the coalescing pair. For
  the gold near-EP the shift is only $5.9\%$, reflecting the
  localisation of the gold doublet. The effective splitting is
  $L$-dependent, growing from $0.179$~eV at $L=4$ to $0.571$~eV at
  $L=12$ (Fig.~\ref{fig:C7}b). The collective enhancement
  renormalises the splitting as well as the biorthogonal overlap.
\end{enumerate}

A caveat for the experimental relevance of the chain-level result deserves
emphasis. The chain analysis is performed at a weakened Rashba amplitude
($\lambda_R^{\rm chain}=0.05$~eV), not the physical value of $\sim 0.6$~eV,
because at the physical value the coherent spin-mixing per bond exceeds the
EP condition $\Delta\gamma/4$ and no near-EP is reachable within the present
model. The chain-level EP should therefore be read as a controlled test of
the collective mechanism, not as a prediction for a specific molecular
junction. 
The physical extensions discussed in Sec.~\ref{sec:chain_setup} (larger
substrate spin polarisation, smaller intra-chain hopping, or Rashba coupling
confined to a subset of bonds) could make the near-EP accessible at physical
Rashba, and the qualitative physics is independent of this caveat.

The results suggest a new class of experimental tests of the spinterface
mechanism: chirality-selective transport, chain-length scaling of
non-Hermitian spectral features, and dressing-dependent residual gaps. The
Petermann factor is directly accessible in transport through shot noise: for
a coherent single-channel conductor the Fano factor $F = S/(2eI)$ satisfies
$F\geq 1/K$~\cite{Schomerus2010}, so $F_{\rm min}(L)\propto L^{-\beta}$
provides a direct signature of the collective scaling without requiring
resolution of individual transmission eigenvalues. None of these tests
requires reaching an exact EP: the near-EP features are already visible at
$K\sim 6.7$ (gold) and $K\sim 135$ (chain).

The spinterface model is therefore a controlled platform for non-Hermitian
physics in molecular transport. The reciprocity breaking that produces the
odd part of the spin current in a magnetically asymmetric device also
produces nonlinear reciprocity breaking at exceptional points, and the two
aspects are complementary views of a single underlying mechanism. We hope
that these results will stimulate further theoretical and experimental work
on the non-Hermitian physics of molecular junctions.

\begin{acknowledgments}
We gratefully acknowledge financial support from Olle Engkvist Stiftelse. 
\end{acknowledgments}

\nocite{*}
\bibliography{ref}

@CONTROL{REVTEX42Control}

@CONTROL{apsrev42Control,author="08",editor="1",pages="0",title="0",year="1"}

@article{Alwan2021,
  title   = {Spinterface Origin for the Chirality-Induced Spin-Selectivity Effect},
  author  = {Alwan, Seif and Dubi, Yonatan},
  journal = {Journal of the American Chemical Society},
  volume  = {143},
  number  = {35},
  pages   = {14235--14241},
  year    = {2021},
  doi     = {10.1021/jacs.1c05637},
  note    = {PMID: 34460257}
}

@article{Ray99,
  title   = {Asymmetric Scattering of Polarized Electrons by Organized Organic Films of Chiral Molecules},
  author  = {Ray, K. and Ananthavel, S. P. and Waldeck, D. H. and Naaman, R.},
  journal = {Science},
  volume  = {283},
  number  = {5403},
  pages   = {814--816},
  year    = {1999},
  doi     = {10.1126/science.283.5403.814}
}

@article{Gohler11,
  title   = {Spin Selectivity in Electron Transmission Through Self-Assembled Monolayers of Double-Stranded {DNA}},
  author  = {G{\"o}hler, B. and Hamelbeck, V. and Markus, T. Z. and Kettner, M. and Hanne, G. F. and Vager, Z. and Naaman, R. and Zacharias, H.},
  journal = {Science},
  volume  = {331},
  number  = {6019},
  pages   = {894--897},
  year    = {2011},
  doi     = {10.1126/science.1199339}
}

@article{Evers22,
  title   = {Theory of Chirality Induced Spin Selectivity: Progress and Challenges},
  author  = {Evers, Ferdinand and Aharony, Amnon and Bar-Gill, Nir and Entin-Wohlman, Ora and Hedeg{\aa}rd, Per and Hod, Oded and Jelinek, Pavel and Kamieniarz, Grzegorz and Lemeshko, Mikhail and Michaeli, Karen and Mujica, Vladimiro and Naaman, Ron and Paltiel, Yossi and Refaely-Abramson, Sivan and Tal, Oren and Thijssen, Joseph and Thoss, Michael and van Ruitenbeek, J. M. and Venkataraman, Latha and Waldeck, David H. and Yan, Binghai and Kronik, Leeor},
  journal = {Advanced Materials},
  volume  = {34},
  number  = {13},
  pages   = {2106629},
  year    = {2022},
  doi     = {10.1002/adma.202106629}
}

@article{Naaman2019,
  title   = {Chirality-Induced Spin Selectivity: A New Phenomenon in Electron Transport},
  author  = {Naaman, Ron and Waldeck, David H.},
  journal = {Annual Review of Physical Chemistry},
  volume  = {70},
  pages   = {1--23},
  year    = {2019},
  doi     = {10.1146/annurev-physchem-050317-021237}
}

@article{Gutierrez2012,
  title   = {Spin-selective transport through helical molecular systems},
  author  = {Gutierrez, R. and D{\'\i}az, E. and Naaman, R. and Cuniberti, G.},
  journal = {Physical Review B},
  volume  = {85},
  number  = {8},
  pages   = {081404(R)},
  year    = {2012},
  doi     = {10.1103/PhysRevB.85.081404}
}

@article{Shitade2020,
  title   = {Geometric spin--orbit coupling and chirality-induced spin selectivity},
  author  = {Shitade, Atsuo and Minamitani, Emi},
  journal = {New Journal of Physics},
  volume  = {22},
  number  = {11},
  pages   = {113023},
  year    = {2020},
  doi     = {10.1088/1367-2630/abc920}
}

@article{Sarkar2025,
    author = {Sarkar, Subhajit and Sharoni, Amos and Monti, Oliver L. A. and Dubi, Yonatan},
    title = {The Spinterface Mechanism for the Chiral-Induced Spin Selectivity
             Effect: A Critical Perspective},
    journal = {ACS Nano},
    volume = {19},
    number = {43},
    pages = {37484-37503},
    year = {2025},
    month = {11},
    issn = {1936-0851},
    doi = {10.1021/acsnano.5c16086},
    url = {https://doi.org/10.1021/acsnano.5c16086},
    eprint = {https://pubs.acs.org/ancac3/article-pdf/19/43/37484/42271696/nn5c16086.pdf},
}

@article{Michaeli2019,
  title   = {Origin of Spin-Dependent Tunneling Through Chiral Molecules},
  author  = {Michaeli, Karen and Naaman, Ron},
  journal = {The Journal of Physical Chemistry C},
  volume  = {123},
  number  = {28},
  pages   = {17043--17048},
  year    = {2019},
  doi     = {10.1021/acs.jpcc.9b05020}
}

@article{Naaman2020,
  title   = {Chiral Molecules and the Spin Selectivity Effect},
  author  = {Naaman, Ron and Paltiel, Yossi and Waldeck, David H.},
  journal = {The Journal of Physical Chemistry Letters},
  volume  = {11},
  number  = {9},
  pages   = {3660--3666},
  year    = {2020},
  doi     = {10.1021/acs.jpclett.0c00474},
  note    = {Explicitly states that spin polarization exceeding 80\% is known experimentally}
}

@article{Kulkarni2020,
  title   = {Highly Efficient and Tunable Filtering of Electrons' Spin by Supramolecular Chirality of Nanofiber-Based Materials},
  author  = {Kulkarni, Chidambar and Mondal, Amit Kumar and Das, Tapan Kumar and Grinbom, Gal and Tassinari, Francesco and Mabesoone, Mathijs F. J. and Meijer, E. W. and Naaman, Ron},
  journal = {Advanced Materials},
  volume  = {32},
  number  = {7},
  pages   = {1904965},
  year    = {2020},
  doi     = {10.1002/adma.201904965},
  note    = {Reports spin filtering in excess of 85\% at room temperature}
}

@article{Das2022,
  title   = {Temperature-Dependent Chiral-Induced Spin Selectivity Effect: Experiments and Theory},
  author  = {Das, Tapan Kumar and Tassinari, Francesco and Naaman, Ron and Fransson, Jonas},
  journal = {The Journal of Physical Chemistry C},
  volume  = {126},
  number  = {6},
  pages   = {3257--3264},
  year    = {2022},
  doi     = {10.1021/acs.jpcc.1c10570},
  note    = {Discusses temperature and bias dependence of CISS polarisation}
}

@article{Sun2025,
  title   = {Robust chirality-induced spin selectivity in topologically chiral molecular knots},
  author  = {Sun, Xi and Zhang, Kai-Yuan and Zhou, Shu-Zheng and Fu, Hua-Hua},
  journal = {Nature Communications},
  volume  = {17},
  pages   = {1231},
  year    = {2025},
  doi     = {10.1038/s41467-025-67988-8}
}

@article{Hossain2025,
  title   = {Observation of Two-Terminal {CISS} Magnetoresistance with Nonmagnetic Contacts},
  author  = {Hossain, Md Anik and Illescas-Lopez, Sara and Firouzeh, Seyedamin and Cuerva, Juan Manuel and Alvarez de Cienfuegos, Luis and Pramanik, Sandipan},
  journal = {Nano Letters},
  volume  = {25},
  number  = {24},
  pages   = {9623--9630},
  year    = {2025},
  doi     = {10.1021/acs.nanolett.5c01297}
}

@article{Yang2020,
  title   = {Detecting chirality in two-terminal electronic devices},
  author  = {Yang, Xu and van der Wal, Caspar H. and van Wees, Bart J.},
  journal = {Physical Review B},
  volume  = {101},
  number  = {11},
  pages   = {115410},
  year    = {2020},
  doi     = {10.1103/PhysRevB.101.115410},
  note    = {Discusses how CISS can be separated from conventional spin-valve magnetoresistance, including chirality-based spin-valve architectures and the distinction between time-reversal-breaking magnetism and spin-orbit CISS}
}

@article{Safari2024,
  title   = {Spin-Selective Electron Transport Through Single Chiral Molecules},
  author  = {Safari, Mohammad Reza and Matthes, Frank and Schneider, Claus M. and Ernst, Karl-Heinz and B{\"u}rgler, Daniel E.},
  journal = {Small},
  volume  = {20},
  number  = {22},
  pages   = {e2308233},
  year    = {2024},
  doi     = {10.1002/smll.202308233}
}

@article{Nguyen2024,
  title   = {The mechanism of the molecular {CISS} effect in chiral nano-junctions},
  author  = {Nguyen, Thi Ngoc Ha and Salvan, Georgeta and Hellwig, Olav and Paltiel, Yossi and Baczewski, Lech Thomasz and Tegenkamp, Christoph},
  journal = {Chemical Science},
  volume  = {15},
  pages   = {14905},
  year    = {2024},
  doi     = {10.1039/D4SC04435E},
  note    = {Argues that the CISS effect is not limited to spinterfaces and can arise from spin-polarised interface effects; relevant for the distinction between chiral and achiral interface contributions}
}

@book{arfken2013mathematical,
  author    = {Arfken, George B. and Weber, Hans J. and Harris, Frank E.},
  title     = {Mathematical Methods for Physicists: A Comprehensive Guide},
  edition   = {7th},
  year      = {2013},
  publisher = {Academic Press},
  isbn      = {978-0-12-384654-9}
}

@book{stefanucci2025nonequilibrium,
  author    = {Stefanucci, Gianluca and van Leeuwen, Robert},
  title     = {Nonequilibrium Many-Body Theory of Quantum Systems: A Modern Introduction},
  edition   = {2nd},
  year      = {2025},
  publisher = {Cambridge University Press},
  isbn      = {978-1-009-42854-9}
}

@article{Zhao2025NHSE,
  author  = {Zhao, Yufei and Zhang, Kai and Xiao, Jiewen and Sun, Kai and Yan, Binghai},
  title   = {Magnetochiral charge pumping due to charge trapping and skin effect in chirality-induced spin selectivity},
  journal = {Nat. Commun.},
  volume  = {16},
  pages   = {37},
  year    = {2025},
  doi     = {10.1038/s41467-024-55433-1}
}

@article{Theiler2026SciAdv,
  author  = {Theiler, Pius M. and Driessen, Sander and Beard, Matthew C.},
  title   = {{$PT$} symmetry enforced twin exchange as the origin of chirality-induced spin selectivity},
  journal = {Science Advances},
  year    = {2026},
  note    = {arXiv:2505.06173}
}

@article{Mayer2026EnantioEP,
  title   = {Enantiosensitive exceptional points in open chiral systems},
  author  = {Mayer, Nicola and L{\"o}hr, Alexander and Moiseyev, Nimrod and Ivanov, Misha and Smirnova, Olga},
  journal = {Physical Review A},
  volume  = {114},
  pages   = {013517},
  year    = {2026},
  doi     = {10.1103/7rtp-3xts}
}

@article{Yang2024SpiroEP,
  title   = {Exceptional Field Effect and Negative Differential Conductance in Spiro-Conjugated Single-Molecule Junctions},
  author  = {Yang, Caiyao and Cao, Jiawen and Lin, Jin-Liang and Wu, Hao and Zhang, Hao-Li and Guo, Xuefeng},
  journal = {Journal of the American Chemical Society},
  volume  = {146},
  number  = {43},
  pages   = {29703--29711},
  year    = {2024},
  doi     = {10.1021/jacs.4c10924}
}

@article{Fransson2023ChiralPhonon,
  author  = {Fransson, Jonas},
  title   = {Chiral phonon induced spin polarization},
  journal = {Phys. Rev. Res.},
  volume  = {5},
  pages   = {L022039},
  year    = {2023},
  doi     = {10.1103/PhysRevResearch.5.L022039}
}

@article{Juraschek2025ChiralPhonons,
  author  = {Juraschek, Dominik M. and Geilhufe, R. Matthias and Zhu, Hanyu and Basini, Martina and Baum, Peter and Baydin, Andrey and Chaudhary, Swati and Fechner, Michael and Flebus, Benedetta and Grissonnanche, Gael and Kirilyuk, Andrei I. and Lemeshko, Mikhail and Maehrlein, Sebastian F. and Mignolet, Maxime and Murakami, Shuichi and Niu, Qian and Nowak, Ulrich and Romao, Carl P. and Rostami, Habib and Satoh, Takuya and Spaldin, Nicola A. and Ueda, Hiroki and Zhang, Lifa},
  title   = {Chiral phonons},
  journal = {Nat. Phys.},
  volume  = {21},
  number  = {10},
  pages   = {1532--1540},
  year    = {2025},
  doi     = {10.1038/s41567-025-03001-9}
}

@article{Fransson2025ChiralPhononsFerro,
  author  = {Fransson, Jonas and Kapon, Yael and Brann, Lior and Yochelis, Shira and Sasselov, Dimitar D. and Paltiel, Yossi and Ozturk, S. Furkan},
  title   = {Chiral Phonons Enhance Ferromagnetism},
  journal = {J. Phys. Chem. Lett.},
  volume  = {16},
  number  = {8},
  pages   = {2001--2007},
  year    = {2025},
  doi     = {10.1021/acs.jpclett.5c00304}
}

@misc{Alwan2026QSpinterface,
      title={Quantum Formulation of the Spinterface Model}, 
      author={Seif Alwan},
      year={2026},
      eprint={2610.28406},
      archivePrefix={arXiv},
      primaryClass={cond-mat.mes-hall},
      url={https://arxiv.org/abs/2610.28406}, 
}

@article{Yang2019PRB,
  title   = {Spin-dependent electron transmission model for chiral molecules in mesoscopic devices},
  author  = {Yang, Xu and van der Wal, Caspar H. and van Wees, Bart J.},
  journal = {Phys. Rev. B},
  volume  = {99},
  pages   = {024418},
  year    = {2019},
  doi     = {10.1103/PhysRevB.99.024418}
}

@article{Geng2023PRB,
  title   = {Nonreciprocal charge and spin transport induced by non-Hermitian skin effect in mesoscopic heterojunctions},
  author  = {Geng, H. and Wei, J. Y. and Zou, M. H. and Sheng, L. and Chen, W. and Xing, D. Y.},
  journal = {Phys. Rev. B},
  volume  = {107},
  pages   = {035402},
  year    = {2023},
  doi     = {10.1103/PhysRevB.107.035402}
}

@article{Solow2025PRB,
  title   = {Signatures of exceptional points in multiterminal superconductor--normal metal junctions},
  author  = {Solow, Oliver and Flensberg, Karsten},
  journal = {Phys. Rev. B},
  volume  = {112},
  pages   = {L161402},
  year    = {2025},
  doi     = {10.1103/PhysRevB.112.L161402}
}

@article{Fransson2020PRB,
  title   = {Vibrational origin of exchange splitting and chiral-induced spin selectivity},
  author  = {Fransson, Jonas},
  journal = {Phys. Rev. B},
  volume  = {102},
  pages   = {235416},
  year    = {2020},
  doi     = {10.1103/PhysRevB.102.235416}
}

@article{Mao2025PRB,
  title   = {Dynamical theory of chiral-induced spin selectivity in electron donor--chiral molecule--acceptor systems},
  author  = {Mao, Y. and Guo, A.-M. and Sun, Q.-F.},
  journal = {Phys. Rev. B},
  volume  = {111},
  pages   = {205417},
  year    = {2025},
  doi     = {10.1103/PhysRevB.111.205417}
}

@article{Zhang2025PRB,
  title   = {Realizing exceptional points of any order in the presence of symmetry},
  author  = {Zhang, Xiang and Hu, Kangyi and Li, Fuxiang},
  journal = {Phys. Rev. B},
  volume  = {112},
  pages   = {224305},
  year    = {2025},
  doi     = {10.1103/PhysRevB.112.224305}
}

@article{ElGanainy2018,
  author  = {El-Ganainy, Ramy and Makris, Konstantinos G. and Khajavikhan, Mercedeh and Musslimani, Ziad H. and Rotter, Stefan and Christodoulides, Demetrios N.},
  title   = {Non-{H}ermitian physics and {PT} symmetry},
  journal = {Nat. Phys.},
  volume  = {14},
  pages   = {11--19},
  year    = {2018},
  doi     = {10.1038/nphys4323}
}

@article{Miri2019,
  author  = {Miri, Mohammad-Ali and Al{\`u}, Andrea},
  title   = {Exceptional points in optics and photonics},
  journal = {Science},
  volume  = {363},
  number  = {6422},
  pages   = {eaar7709},
  year    = {2019},
  doi     = {10.1126/science.aar7709}
}

@article{Bergholz2024,
  author  = {Bergholtz, Emil J. and Budich, Jan Carl and Kunst, Flore K.},
  title   = {Exceptional topology of non-{H}ermitian systems},
  journal = {Rev. Mod. Phys.},
  volume  = {93},
  pages   = {015005},
  year    = {2021},
  doi     = {10.1103/RevModPhys.93.015005}
}

@article{Petermann1979,
  author  = {Petermann, Klaus},
  title   = {Calculated spontaneous emission factor for double-heterostructure injection lasers with gain-induced waveguiding},
  journal = {IEEE J. Quantum Electron.},
  volume  = {15},
  number  = {7},
  pages   = {566--570},
  year    = {1979},
  doi     = {10.1109/JQE.1979.1070064}
}

@article{Siegman1989,
  author  = {Siegman, A. E.},
  title   = {Excess spontaneous emission in non-{H}ermitian optical systems. {I}. {L}aser amplifiers},
  journal = {Phys. Rev. A},
  volume  = {39},
  pages   = {1253--1263},
  year    = {1989},
  doi     = {10.1103/PhysRevA.39.1253}
}

@article{Schomerus2010,
  author  = {Schomerus, Henning},
  title   = {Quantum noise and self-sustained radiation of {PT}-symmetric systems},
  journal = {Phys. Rev. Lett.},
  volume  = {104},
  number  = {23},
  pages   = {233601},
  year    = {2010},
  doi     = {10.1103/PhysRevLett.104.233601}
}

@article{Naaman2015,
   author = "Naaman, Ron and Waldeck, David H.",
   title = "Spintronics and Chirality: Spin Selectivity in Electron Transport Through Chiral Molecules", 
   journal= "Annual Review of Physical Chemistry",
   year = "2015",
   volume = "66",
   number = "Volume 66, 2015",
   pages = "263-281",
   doi = "https://doi.org/10.1146/annurev-physchem-040214-121554",
   url = "https://www.annualreviews.org/content/journals/10.1146/annurev-physchem-040214-121554",
   publisher = "Annual Reviews",
   issn = "1545-1593",
   type = "Journal Article",
  }

@article{Bloom2024ChemRev,
  author  = {Bloom, Brian P. and Paltiel, Yossi and Naaman, Ron and Waldeck, David H.},
  title   = {Chiral Induced Spin Selectivity},
  journal = {Chemical Reviews},
  volume  = {124},
  number  = {4},
  pages   = {1950--1991},
  year    = {2024},
  doi     = {10.1021/acs.chemrev.3c00661}
}

\end{document}